\documentclass[letterpaper,11pt,abstracton]{scrartcl}

\RequirePackage{fullpage}

\RequirePackage{enumerate}
\RequirePackage[latin1]{inputenc}
\RequirePackage{amsmath}
\RequirePackage{amssymb}

\usepackage{algorithmicx}
\usepackage{algorithm}
\usepackage[noend]{algpseudocode}

\usepackage{amsmath,amsfonts}
\usepackage{epsfig}

\newcommand{\qed}{\hfill$\diamond$}
\newcommand{\pf}{{\textbf Proof: }}
\newtheorem{theorem}{Theorem}[section]

\newtheorem{lemma}[theorem]{Lemma}
\newtheorem{tm}[theorem]{Theorem}

\newtheorem{obs}[theorem]{Observation}
\newtheorem{cor}[theorem]{Corollary}

\newtheorem{dfn}[theorem]{Definition}

\newcommand{\beq}{\begin{equation}}
\newcommand{\eeq}{\end{equation}}

\begin{document}

\title{Ordering with precedence constraints and budget minimization}

\author{Akbar Rafiey \thanks{Simon Fraser University email : arafiey@sfu.ca} \and 
Jeff Kinne \thanks{Indiana State University email : jkinne@cs.indstate.edu Supported by Indiana State University COMPETE grant} \and
J\'an Ma\v nuch \thanks{University of British Columbia email : jmanuch@cs.ubc.ca  Supported by NSERC Canada} \and
Arash Rafiey \thanks{Indiana State University and Simon Fraser University email :  
arash.rafiey@indstate.edu Supported by Indiana State University COMPETE grant} \footnote{Names of the last three authors appear alphabetically}
}

\date{}
\maketitle

\begin{abstract}
We introduce a variation of the scheduling with precedence
constraints problem that has applications to molecular folding and
production management. We are given a bipartite graph
$H=(B,S)$.  Vertices in $B$ are thought of as goods or services that must be
\emph{bought} to produce items in $S$ that are to be \emph{sold}.  An
edge from $j\in S$ to $i\in B$ indicates that the production of
$j$ requires the purchase of $i$.  Each vertex in $B$ has a cost, and
each vertex in $S$ results in some gain.  The goal
is to obtain an ordering of $B\cup S$ that respects the precedence
constraints and maximizes the minimal net profit encountered as the vertices
are processed.  We call this optimal value the \emph{budget} or
\emph{capital} investment required for the bipartite graph, and refer to our
problem as \emph{the bipartite graph ordering problem}.

The problem is equivalent to a version of an NP-complete molecular folding problem
that has been studied recently \cite{MTSC09}. Work on
the molecular folding problem has focused on heuristic algorithms and
exponential-time exact algorithms for the un-weighted problem where costs
are $\pm 1$ and when restricted to graphs arising from RNA folding.

The present work seeks exact algorithms for solving the bipartite
ordering problem.  We demonstrate an algorithm that computes
the optimal ordering in time $O^*(2^n)$ when $n$ is the number of vertices
in the input bipartite graph. We give non-trivial polynomial time algorithms for finding
the optimal solutions for bipartite permutation graphs, trivially perfect bipartite graphs, co-bipartite graphs.

We introduce a general strategy that can
be used to find an optimal ordering in polynomial time for bipartite graphs
that satisfy certain properties. One of our ultimate goals is to completely characterize
the classes of graphs for which the problem can be solved exactly
in polynomial time.

%We develop a technique to obtain a polynomial-time solution to the bipartite graph
%ordering problem for convex bipartite graphs.
% to  bipartite graph
% ordering problem for bipartite graphs
%
% convext biparite graph to find the optimal solution
%
% Our main result is a general strategy that can
% be used to find an optimal ordering in polynomial time for bipartite graphs
% that satisfy certain properties.  We apply the technique to a variety
% of graph classes, obtaining polynomial-time solutions to the bipartite graph
% ordering problem for bipartite graphs that are convex, trivially perfect,
% co-bipartite graphs, and trees.

%At the conclusion we show our strategy works for a general class of bipartite graphs where the white vertices are the vertices of
%a tree with bounded number of leaves and each black vertex is adjacent to a sub-tree of this tree. We also propose several
%open problems which are of interest both for theoretical reasons and
%in practice.
\end{abstract}

\section{Motivation and Introduction}\label{sec:intro}

%Arash
\paragraph{Job Scheduling with Precedence Constraints}
The setting of job scheduling with precedence constraints is a natural
one that has been much studied (see, e.g., \cite{CS05,MMS04}).  A number of
variations of the problem have been studied; we begin by stating one.
The problem is formulated as a directed acyclic graph where
the vertices are jobs and arcs between the vertices impose precedence
constraints.  Job $j$ must be executed after job $i$ is completed if
there is an arc from $j$ to $i$.  Each job $i$ has a weight $w_i$ and
processing time $t_i$.  A given ordering of executing the jobs results
in a completion time $C_i$ for each job.  Previous work has focused on
minimizing the weighted completion time $\sum_{i=1}^{i=n}w_iC_i$.
This can be done in the single-processor or multi-processor setting,
and can be considered in settings where the precedence graph is from
a restricted graph class.  The general problem of finding an ordering that
respects the precedence constraints and minimizes the weighted
completion time is NP-complete.  Both approximation algorithms and
hardness of approximation results are known \cite{AMMS11,AMS07,MMS04,woginger1}.

%
%
%  and the goal is to find an ordering of the jobs
% that respects the constraint and minimizes
% where $C_i$ is the completion time of job $i$.  Previous work has
% focused on minimizing the
%
% A sub-case of the problem is when the constraint
% digraph is bipartite and all arcs (precedence) are from left to right
% \cite{woginger1} and the weight of the jobs in the left side are one
% and their processing time is zero and the weight of the jobs in the
% right side are zero but their processing time is one.
%
% Each job $j$ has a net profit (positive or negative) $p_j$.

\paragraph{Our Problem -- Optimizing the Budget}
In the present work, we consider a different objective than previous
works.  In our setting, each job $j$ has a net profit (positive or
negative) $p_j$.  Our focus is on the \emph{budget} required to
realize a given ordering or schedule, and we disregard the processing
time.  We imagine that the jobs are divided between those with negative
$p_i$, jobs $B$ that must be \emph{bought}, and jobs with a non-negative
$p_i$, jobs $S$ that are \emph{sold}.  $B$ could consist of raw inputs
that must be purchased in bulk in order to produce goods $S$ that
can be sold.  A directed graph
$H=(B,S)$ encodes the precedence constraints inherent in the
production: an arc from $j\in S$ to $i\in B$ implies that item $i$
must be bought before item $j$ can be produced and sold.
At each step $1 \le r \le n$ of the process, let $j_1, j_2, ..., j_r$ be the jobs processed
thus far, and let $bg_r = \sum_{i=1}^r p_{j_i}$ be the total budget
up to this point.  Our goal is an ordering that respects the precedence
constraints and keeps the minimal value of $bg_r$ as high as possible.
One can view (the absolute value of) this optimal value as the \emph{capital} investment required
to realize the production schedule.

%INSERT REFERENCE TO FIGURE THAT HAS
%AN EXAMPLE, GIVE A POSSIBLE ORDERING AND SHOW HOW $bg_r$ CHANGES.

In this work we assume $H$ is a bipartite graph with all arcs from
$S$ to $B$.  This models the situation where each item to be produced and
sold depends on certain inputs that must be purchased.
We call this the problem of \emph{ordering with precedence
  constraints and budget minimization on bipartite graphs} but refer
to the problem as the \emph{bipartite graph ordering problem}.

\paragraph{Applications}
The bipartite graph ordering problem is a natural variation of scheduling with
precedence constraints problems.  As described above the problem can be used
to model the purchase of supplies and production of goods when purchasing in
bulk.  Another way to view the problem is that the items in $B$ are training sessions that employees must complete before employees (vertices in $S$) can
begin to work.

\begin{figure}[htbp]
\begin{center}
\includegraphics[scale=0.5]{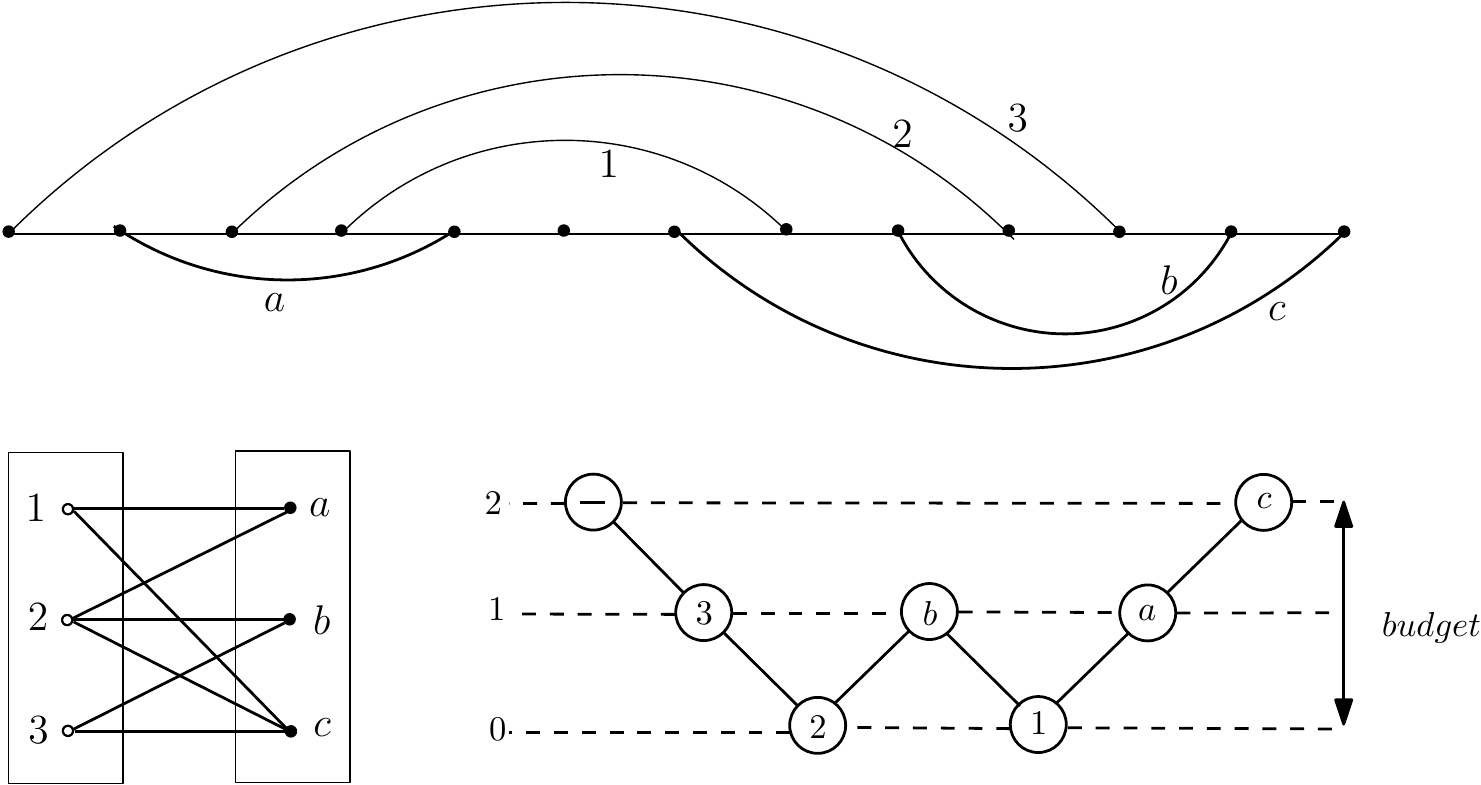}
\caption{
\footnotesize{
  The top graph is an instance of the RNA folding problem, with folds $1$, $2$,
  and $3$ to be removed (bought), folds $a$, $b$, and $c$ to be added (sold); an edge cannot
  be added until edges that cross it are removed. A budget of two is needed and an optimal ordering is $3$, $2$, $b$, $1$, $a$, $c$.
}
}
\label{fig:1}
\end{center}
\end{figure}

We began studying the problem as a generalization of an optimization
problem in molecular folding.  The folding problem asks for the energy
required for secondary RNA structures to be transformed from a given
initial folding configuration $\mathcal{C}_1$ into a given final folding
configuration $\mathcal{C}_2$ \cite{GFWT08,MH98,TMRM09}. The bipartite graph ordering problem models
this situation as follows: vertices in $B$ are folds that are to be
removed from $\mathcal{C}_1$, vertices in $S$ are folds that are to be
added, and an edge from $j$ to $i$ indicates that fold $i$ must be
removed before fold $j$ can be added.  The price $p_i$ of a vertex is set
according to the net energy that would result from allowing the given
fold to occur, with folds that must be broken requiring a positive
energy and folds that are to be added given a negative energy.  The
goal is to determine a sequence of transformations that respects these
constraints and still keeps the net energy throughout at a minimum
\footnote{Note that the molecular folding problem is a minimization problem,
  and can be made a maximization problem by negating the energies.
}.
Figure \ref{fig:1} shows how an instance of the RNA folding problem is
transformed into the bipartite graph ordering problem.

\paragraph{Previous Work}
The molecular folding problem has been studied only in the
setting of unit prices and most attention has been devoted to graph classes corresponding to typical
folding patterns (in particular for so-called circle bipartite graphs).
% \cite{MH98,FHMS01,FHSW02,GFWT08,TMRM09}
\cite{MTSC09} shows that the molecular folding problem is NP-complete
even when restricted to circle bipartite graphs; thus the bipartite graph ordering
problem is NP-complete as well when restricted to circle bipartite graphs
\footnote{ A graph $G$ is called a \emph{circle graph} if the vertices
  are the chords of a circle and two vertices are adjacent if their
  chords intersect.  The circle bipartite graphs can be represented as two
  sets $A,B$ where the vertices in $A$ are a set of non-crossing arcs
  on a real line and the vertices in $B$ are a set of non-crossing
  arcs from a real line; there is an edge between a vertex in $A$ and
  a vertex in $B$ if their arcs cross.  The top graph in Figure
  \ref{fig:1} is a circle bipartite graph shown with this representation.}.

Previous work on the folding problem has focused on exact algorithms that
take exponential time and on heuristic algorithms \cite{FHMS01}.

% Akbar

%\vspace{-.3in}

There has been considerable study of scheduling with precedence constraints,
but to our knowledge there has not been any work by that community on the
objective function we propose (budget minimization).

\subsection{Our Results}

We introduce the bipartite graph ordering problem, which is equivalent to a
generalization of a molecular folding problem.  We initiate the study
of which graph classes admit polynomial-time exact solutions. %and improve on the
%previous best-known exact algorithm for arbitrary graphs.
We also
give the first results for the weighted version of the problem; previous
work on the molecular folding problem assumed unit costs for all folds.

\paragraph{Exponential-time Exact Algorithm}
We first give an exact algorithm for arbitrary bipartite graphs.
\begin{tm}\label{thm:exp}
  Given a bipartite graph $H=(B,S)$, the bipartite graph ordering problem on $H$ can be
  solved in (a) time and space $O^*(2^n)$, and (b) time $O^*(4^n)$ and polynomial
  space, where $n=|B\cup S|$.
\end{tm}
The previous best exact algorithm for the molecular folding
problem on circle bipartite graphs has running time $n^{O(K)}$, where $K$ is the optimal budget
\cite{TMRM09}.

We observe that $K$ can be $\Omega(n)$ when vertex prices
are $\pm 1$ (and can be much larger when vertex prices can be arbitrary), as follows.
%, Appendix part B. In the interest of structural graph theory, one can see that there are bipartite graphs for which the required budget is linear in terms of number of vertices.
Let $\mathcal{P}$ be a projective plane of order $p^2+p+1$ with $p$ prime.  The projective plane of order $n=p^2+p+1$ consists of $n$ lines each consisting of
precisely $p+1$ points, and $n$ points which each are intersected by precisely $p+1$
lines.  We construct a bipartite graph with each vertex in $B$ corresponding
to a line from the projective plane, each vertex in $S$ corresponding
to a point from the projective plane, and a connection from $b\in B$ to
$s\in S$ if the projective plane point corresponding to $s$ is contained in
the line corresponding to $B$.  Vertices in $B$ are given weight -1, and
vertices in $S$ are given weight 1.
Note that the degree of each vertex in $B$ is $p+1$. One can observe that the neighbourhood of every set of $p+1$ vertices in $S$ is at
least $p^2- { p \choose 2 }$. This implies
that in order to be able to sell the first $p+1$ vertices in $S$ the budget decreases by at least $p^2- { p \choose 2 }+p$.

\paragraph{Polynomial-time Cases}
%Our main result gives a general technique that can be used to solve
%the bipartite graph ordering problem for classes of bipartite graphs that satisfy
%certain properties.

We develop algorithms for solving  a number of bipartite graph classes.
These bipartite graph classes are briefly defined after the theorem statement and
discussed further in Sections \ref{sec:poly-TP-CB} and \ref{sec:poly-permutation}.

\begin{tm}\label{thm:main}
  Given a bipartite graph $H=(B,S)$, the bipartite graph ordering problem on $H$ can be
  solved in polynomial time if $H$ is one of the following:
  a bipartite permutation graph, a trivially perfect bipartite graph, a co-bipartite graph or a tree.
\end{tm}

The bipartite graphs we consider here have been considered for other types of optimization problems.
In particular \emph{bipartite permutation graphs} also known as proper interval bipartite graphs
(those for which there exists an ordering of the vertices in $B$ where the neighborhood of each vertex in $S$ is a set of consecutive vertices (interval) and the intervals can be chosen
so that they are inclusion free) are of interest in graph homomorphism problems \cite{GHRY08} and also
%in biology \cite{AKNW95,LH03}
in energy production applications where resources (in our case bought vertices) can be assigned (bought) and used (sold) within a number of
successive time steps \cite{KKRS13,MS12}.
There are recognition algorithms for bipartite permutation graphs \cite{GHRY08,SBS98}.
A bipartite graph is called \emph{trivially perfect} if it is obtained from a union of two trivially perfect bipartite graphs $H_1,H_2$ or by joining every sold vertex in trivially perfect bipartite graph
$H_1$ to every bought vertex in trivially perfect bipartite graph $H_2$. A single vertex is also a trivially perfect bipartite graph.
These bipartite graphs have been considered in \cite{CEM13,EKLT12,MMS04}.  \emph{Co-bipartite graphs} have a similar definition
with a slightly different join operation. See Section \ref{sec:poly-TP-CB} for the precise definitions.

% JJK, updated this paragraph ...
For trivially perfect bipartite graphs and co-bipartite graphs, due to the recursive nature
of the definition of these graphs it is natural to attempt a divide and conquer strategy.
However, a simple approach of solving sub-problems and using these to build up to a solution of the whole
problem fails because one may need to consider all possible orderings of combining the sub problems.

In section \ref{sec:general-strategy} we develop a general approach that can be applied to the graph classes
mentioned.
%The main difficulty to overcome in proving Theorem \ref{thm:main} is to determine which
%subgraph to process first, and we develop a general approach that applies to the graph classes
%mentioned.  We frame our algorithm in a way amenable to extending the result to include other
%graph classes, with the hope of fully characterizing the graphs for which our problem is
%polynomially solvable.  We observe in Section \ref{sec:extension} that the algorithm applies
%to a class of graphs more general than convex graphs.

%{\textbf I added } One would think these results are simple as we deal with known and restricted graph classes. But we had to define complex notions for our algorithm as it turns out despite the simple structure of these graph classes, the optimal solution can not be obtained by simply splitting the problem into sub-problems and solving each sub-problem independently and then merging the solution of the sub-problems in a simple manner.
%Also note that a full classification (for which bipartite graphs the optimal budget can be found in polynomial ? ) requires much more work and I hope the reader does not expect us to obtain it in this manuscript. However, we discuss how to apply our strategy to a general class of bipartite graphs where the bought vertices $B$ are the vertices of a tree with a bounded number of leaves and each sold vertex in $S$ is adjacent to a connected sub-tree of this tree.

\paragraph{Arbitrary Vertex Weights}
Each of our results holds where the weights on vertices can be
arbitrary (not only $\pm 1$ as considered by previous work on the
molecular folding problem) except for trees. 

\section{Some Simple Classes of bipartite graphs}\label{sec:simple}
In this section we state some simple facts about the bipartite graph ordering
problem and give a simple self-contained proof that the problem can be
solved for trees.  We provide this section to assist the reader in
developing an intuition for the problem.

\paragraph{\textbf{Bicliques}}
First we note that if $H$ is a biclique with $|B|=K$ then $bg(H)$ (the
budget required to process $H$) is $K$.

As a next step, consider a disjoint union of bicliques $H_1, H_2, ...,
H_m$ where each $H_i$ is a biclique between bought vertices $B_i$ and
sold vertices $S_i$.  Intuition suggests that we should first process
those $H_i$ such that $|S_i| \geq |B_i|$.  This is indeed correct and
is formalized in Lemma \ref{lm:minpos} in Section \ref{sec:main} (the
reader is encouraged to take this intuition for granted while
initially reading the present section).  After processing $H_i$ with
$|S_i|\geq|B_i|$, which we call \emph{positive} (formally defined in
generality in Section \ref{sec:main}), we are left with bicliques
$H_i=(B_i,S_i)$ where $|B_i|>|S_i|$. Up to this point we may have
built up some positive budget.

In processing the remaining $H_i$ the budget steadily goes down --
because the $H_i$ are bicliques and disjoint, and the remaining sets
are not positive.  As we shall see momentarily, we should process those $H_i$ with largest $|S_i|$
first. Suppose on the contrary that $|S_i|>|S_j|$ but an optimal
strategy $opt$ processes $H_j$ right before $H_i$.  If $K$ is the
budget before this step we first have that $K-|B_j|+|S_j| \ge |B_i|$
because otherwise there would not be sufficient budget after
processing $H_j$ to process $H_i$.  Since we assumed that $|S_i| >
|S_j|$ we have $K-|B_i|+|S_i| \ge |B_j|$. Thus, we could first process
$H_i$ and then $H_j$.  We have thus given a method to compute an
optimal strategy for a disjoint union of bicliques: first process
positive sets, and then process bicliques in decreasing order of
$|S_i|$.

\paragraph{\textbf{Paths and Cycles}}
We next consider a few even easier cases.  Note that a simple
path can be processed with a budget of at most 2, and a simple
cycle can be processed with a budget of $2$.

\paragraph{\textbf{Trees and Forests}}
Next we assume the input graph is a tree and the weights are $-1,1$
(for vertices in $B$ and $S$, respectively).  Let $H$ be a tree, or in
general a forest. Note that any leaf has a single neighbor (or none,
if it is an isolated vertex).  We can thus immediately process any
sold leaf $s$ by processing its parent in the tree and then processing
$s$.  This requires an initial budget of only 1.  After repeating the
process to process all sold leaves in $S$, we are left with a forest
where all leaves are bought vertices in $B$.  We can first remove from
consideration any disconnected bought vertices in $B$ (these can, without
loss of generality, be processed last).
% Jeff 14 Jul 15 - please check that the last phrase is correct.  The way
%  we have defined the problem, we should process all of $B$ and not just ignore
%  isolated vertices.  I think we can just process them last.
We are left with a forest $H'$.

We next take a sold vertex $s_1$ (which is not a leaf because all sold
leaves in $S$ have already been processed) and process all of its neighbours.  After processing
$s_1$ we can process $s_1$ and return 1 unit to the budget.  Note that because
$H'$ is a forest, the neighbourhood of $s_1$ has intersection at most 1 with the
neighbourhood of any other sold vertex in $S$.  Because we have already processed all
sold leaves from $H$, we know that only $s_1$ can be processed after processing
its neighbours.

After processing $s_1$, we may be left with some sold leaves in $S$.  If so,
we deal with these as above.  We note that if removing the neighbourhood of
$s_1$ does create any sold leaves, then each of these has at least one
bought vertex in $B$ that is its neighbour and is not the neighbour of any of the other
sold leaves in $S$.  When no sold leaves remain, we pick a
sold vertex $s_2$ and deal with it as we did $s_1$.

This process is repeated until all of $H'$ is processed.  We note that after initially
dealing with all sold leaves in $S$, we gain at most a single sold leaf at a time.
That is, the budget initially increases as we process sold vertices and process their
parents in the tree, and then the budget goes down progressively, only ever
temporarily going up by a single unit each time a sold vertex is processed.
Note that the budget initially increases, and then once it is decreasing
only a single sold vertex is processed at a time.  This implies that the budget required
for our strategy is $|B|-|S|+1$, the best possible budget for a graph with
$1, -1$ weights.
% Jeff 14 Jul 15 - updated this section.  It was referring to some
% definitions and concepts that don't come until later.

% Jeff 14 Jul 15 - updated this section.  It was referring to some
% definitions and concepts that don't come until later.

%Akbar
\section{An Exponential-time Exact Algorithm}\label{sec:exp}
In this section we prove Theorem \ref{thm:exp}.

The authors in \cite{BFKKT} show that any vertex ordering problem on
graphs of a certain form can be solved in both (a) time and space
$O^*(2^n)$, and (b) time $O^*(4^n)$ and polynomial space, where $n$ is
the number of vertices in the graph and $O^*(f(n))$ is shorthand for
$O(f(n)\cdot \textnormal(poly)(n))$.  We show that the ordering
problem can be seen to have the form needed to apply this result.

A vertex ordering on graph $H=(B,S)$ is a bijection $\pi: B\cup
S\rightarrow \lbrace1,2,\cdots, |B\cup S|\rbrace$.
Note that orderings we consider here respect the precedence
constraints given by edges of bipartite graph $H$.
For a vertex
ordering $\pi$ and $v\in B\cup S$, we denote by $\pi_{\prec,v}$ the
set of vertices that appear before $v$ in the ordering.  More
precisely, $\pi_{\prec,v}=\lbrace u\in B\cup S| \pi(u) <
\pi(v)\rbrace$.

Let $\Pi(Q)$ be the set of all permutations of a set $Q$ and $f$ be a
function that maps each couple consisting of a graph $H=(B,S)$ and a
vertex set $Q\subseteq (B\cup S)$ to an integer as
follows:

\begin{center}
$f(H, Q)=|Q\cap S|-| Q\cap B|$.
\end{center}
Note that the function $f$ is polynomially computable. Now, if we
restrict the weights of vertices to be $\pm 1$ (vertices in $B$ have
weight -1 and vertices in $S$ have weight 1) we can express the
bipartite graph ordering problem as follows:

\begin{center}
$bg(H)=\min\limits_{\pi\in\Pi(B\cup S)} \max\limits_{v\in(B\cup S)}f(H, \pi_{\prec,v})$.
\end{center}

The right hand side of this equation is the form required to apply the result of
\cite{BFKKT}, proving Theorem \ref{thm:exp} for the case of $\pm 1$ weights.
The result for arbitrary weights $p_i$, with $p_x$ negative for $x\in B$ and
$p_y$ non-negative for $y\in S$, follows by modifying $f(H,Q)$ to be
$\sum_{y\in Q\cap S} p_y -\sum_{x\in Q\cap B} p_x $.

\section{Definitions and Concepts} \label{sec:main}

\begin{figure}[htbp]
\begin{center}
\includegraphics[scale=0.4]{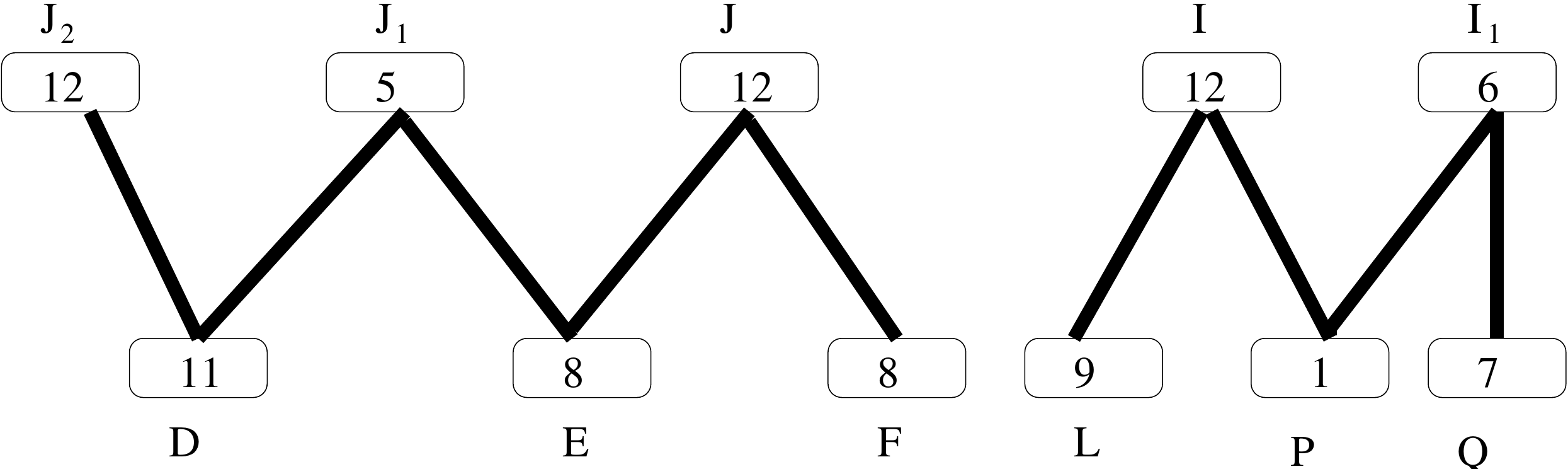}
\caption{ A bipartite permutation graph that we use as an example for
    the definitions related to our algorithm.  Each bold line shows a
    complete connection, i.e. the induced sub-graph by $I \cup L$ is a biclique. The numbers in
    the boxes are the number of vertices.  The sets $J_1, J_2, J, I, I_1$
    are the items $B$ to be bought, with each vertex having weight
    -1.  The sets $D, E, F, L, P, O$ are the items $S$ to be sold,
    with each vertex having weight 1. }
% Jeff 13 Jul 15 - put in mention of current budget for cl and Superset above.
\label{fig:2}
\end{center}
\end{figure}

In this section we define key terms and concepts that are relevant to
algorithms that solve the bipartite graph ordering problem on general bipartite graph.  We use the graph in Figure \ref{fig:2} as an example to
demonstrate each of our definitions.  The reader is encouraged to
consult the figure while reading this section.  %The bipartite graph ordering problem was defined in Section \ref{sec:intro}.
Recall that bipartite graph
$H=(B,S,E)$ encodes the precedence constraints inherent in the
production: an arc from $j\in S$ to $i\in B$ implies that item $i$
must be bought before item $j$ can be produced and sold.
At each step $1 \le r \le n$ of the process, let $j_1, j_2, ..., j_r$ be the jobs processed
thus far, and let $bg_r = \sum_{i=1}^r p_{j_i}$ be the total budget used
up to this point.  Our goal is an ordering that respects the precedence
constraints and keeps the maximal value of $bg_r$ as small as possible.
%One can view (the absolute value of) this optimal value as the \emph{capital} investment required
%to realize the production schedule. for a given
%graph $H$, $bg(H)$ is the optimal budget required to process the graph.

Let $I$ be a set of vertices. $|I|$ refers to the cardinality of set
$I$.  When applying our arguments to weighted graphs, with vertex $x$
having price $p_x$, we let $|I|$ to be $\sum_{x\in I} |p_x|$.
Each of our results holds for weighted graphs by letting $|I|$ refer
to the weighted sum of prices of vertices in $I$ in all definitions and
arguments.

We use $K$ to denote the budget or capital available to process an
input bipartite graph.  As vertices are processed, we let $K$ denote the current
amount of capital available for the rest of the graph.

\begin{dfn}\label{def:Nstar}
  Let $H=(B,S,E)$ be a bipartite graph. For a subset $I\subseteq B$ of bought vertices in $H$,
  let {$N^*(I)$} be the set of all vertices in $S$ whose entire neighborhood lie in $I$.
\end{dfn}

\begin{dfn}\label{def:prime}
  We say a set $I \subseteq B $ is {\em prime} if $N^*(I)$ is non-empty and
  for every proper subset $I'\subset I$, $N^*(I')$ is empty.
\end{dfn}

Note that the bipartite graph induced by a prime set $I$ and $N^*(I)$ is a bipartite clique.
For any strategy to process an input bipartite graph $H$, we look at the
budget at each step of the algorithm. Suppose our initial
budget is $K$. Knowing which subsets of $B$ are prime, one can see that every optimal strategy can be 
modified to start with processing a prime subset 
(Lemma \ref{first-prime}). This
leaves a budget of $K-|I|+|N^*(I)|$ to process the rest of the bipartite graph.
An example for prime sets is given in Figure \ref{fig:2}. For the given graph prime sets are $\{J_1,J_2\}$,
 $J$, $I$, $I_1$ with $N^*(\{J_1,J_2\})=D$, $N^*(J)=F$, $N^*(I)=L$, and $N^*(I_1)=Q$.

\begin{lemma}\label{first-prime}
There is an optimal strategy for \textsc{Bipartite Ordering Problem} on bipartite graph $H=(B,S,E)$ without isolated vertex, 
that starts with a prime set.
\end{lemma}
\pf Let $\pi$ be an optimal strategy that does not start with a prime set. Suppose $2\leq i\leq n$ is the first position in $\pi$ where $\pi(i)\in S$ and
$M=\{\pi(1),\pi(2),\dots,$ $\pi(i-1)\}$. Let set $I\subseteq M$ be the smallest set with $N^*(I)\neq\emptyset$. Note that such a set $I$ exists since all the adjacent vertices to $\pi(i)$ are among vertices in $M$. Observe that changing the processing order on vertices in $M$ does not harm optimality. Therefore, we can change $\pi$ by processing vertices in $I$ at first, without changing the budget. In addition, we can process $N^*(I)$ immediately after processing $I$.
\qed \\

Our algorithm will generally try to first process subsets $I$ that
increase (or at least, do not decrease) the budget.  We call such
subsets \emph{positive}, and call $I$ \emph{negative} if processing it
would decrease the budget.

\begin{dfn}\label{def:budget-for-prime}
  A budget of $I \subset B$ is the minimum budget $r$ needed to process $H[I \cup N^*(I)]$, denoted by  $bg(H[I\cup N^*(I)])=r$.
  For simplicity we write $bg(I)=r$ if $H$ is clear from the context.
\end{dfn}

\begin{dfn}\label{def:positive} A set
  $I \subseteq B$ is called \emph{positive} if $\vert I\vert \leq \vert
  N^{*}(I)\vert$ and it is \emph{negative} if $\vert I\vert > \vert
  N^{*}(I)\vert$.
  For a given budget $K$, $I$ is called \emph{positive minimal} (with respect to budget $K$) if it is positive, $I$ has budget at most
  $K$,  and every other positive  subset of $I$ has budget more than $K$. In
  other words, $I$ is  smallest among all the subsets of $I$ that is positive and has
  budget at most $K$.
\end{dfn}

%{\textbf A budget of $I \subset B$ is actually the minimum capital needed to process $I \cup N^*(I)$ }
For the given graph in Figure \ref{fig:2}, $I_1$ is the only positive minimal set and $N^*(I_1)=O$ contains $7$ vertices. % (indeed $I_1$ is a positive prime set).
Note that, in general, there can be more than one positive minimal set.
Positive minimal sets are key in our algorithms for computing the budget
because these are precisely the sets that we can process
first, as can be seen from Lemma \ref{lm:minpos}.  In the graph of Figure \ref{fig:2}, the positive set $I_1$
would be the first to be processed.

\begin{lemma} \label{lm:minpos}
  Let $H=(B,S,E)$ be a bipartite graph that can be processed with budget at most $K$.
  %Suppose $H$ contains a positive minimal set $I\subset B$ (with respect to budget $K$).
  If $H$ contains a positive minimal set (with respect to $K$) then there is a strategy for $H$ with budget $ K$ 
  that begins by processing a positive minimal
  subset $I$ such that for all $I'\subset I$ we have $|N^*(I')|-|I'|\leq |N^*(I)|-|I|$.
\end{lemma}
% JJK, what about if some parts of S are also processed before I is done being
%  processed?  Is it worth mentioning this?
\pf Let $I$ be a positive minimal set in $H$. Suppose the optimal process $opt$ does not process $I$ all together and hence processes the sequence
$L_1,I_1,L_2,I_2,\dots, L_t,I_t,L_{t+1}$ of disjoint subsets of $B$ where
$I= I_1 \cup I_2 \cup ... \cup I_t$  is a positive minimal set and $L_j \ne \emptyset$, $2 \le j \le t$. Note that according to $opt$
for all $L_i$, $1 \le i \le t+1$ we have $bg(L_i) \le K-|S_2|+|N^*(S_2)|$ in graph $H\setminus S_2$ 
where $S_2= \cup_{j=1}^{i-1}L_j \cup  \cup_{j=1}^{i-1} I_j$.
First consider the case when $|N^*(\cup_{j=1}^{i-1} I_j)|-|\cup_{j=1}^{i-1} I_j|\leq |N^*(I)|-|I|$. 
Let $S_1=\cup_{j=1}^{i-1} L_j$. 
For this case we have
\begin{align}\nonumber
K-|S_2|+|N^*(S_2)|&= K- |S_1|-|\cup_{j=1}^{i-1}I_j|+|N^*(S_2)|\\ \nonumber
				  &=K-|S_1|+|N^*(S_1)|-|\cup_{j=1}^{i-1}I_j|+|N^*   					(\cup_{j=1}^{i-1}I_j)|+ \\ \nonumber
                  &|N^*(S_2)\setminus (N^*(S_1)\cup N^*(\cup_{j=1}^{i-1}I_j))|\\ \nonumber
				  & \leq  K-|S_1|+|N^*(S_1)|-|I|+|N^*(I)|+ \\ \nonumber
                  &|N^*(S_2)\setminus (N^*(S_1)\cup N^*(\cup_{j=1}^{i-1}I_j))|\\ \nonumber
                  & \leq  K-|I\cup S_1|+|N^*(I\cup S_1)|
\end{align}
%where $\alpha=|N^*(S_2)-N^*(S_1)-N^*(\cup_{j=1}^{i-1}I_j)|$.
Therefore $bg(L_i)$ in graph $H_1= H \setminus (I \cup S_1  \cup  N^*(I \cup S_1))$ is at most $K- |I \cup S_1|+|N^*(I \cup S_1)|$.
Together with $bg(I) \le K$, we conclude that, there is another optimal process that considers $I$ 
first and then $L_1,L_2,\dots, L_t,L_{t+1}$ next and then following $opt$. 

Now consider the case when $|N^*(\cup_{j=1}^{i-1} I_j)|-|\cup_{j=1}^{i-1} I_j|\geq |N^*(I)|-|I|$. 
Note that since $I$ is a positive minimal set then processing $H[\cup_{j=1}^{i-1} I_j\cup N^*(\cup_{j=1}^{i-1} I_j)]$ 
needs budget more than $K$ as otherwise $\cup_{j=1}^{i-1} I_j$ contradicts the minimality of $I$. On the other hand, $opt$ processes $H[S_2]$ 
with budget at most $K$. Therefore, during processing $H[S_2]$ there exists a $1\leq \beta\leq i-1$ such 
that $\cup_{j=1}^{\beta} L_j \cup \cup_{j=1}^{\beta} I_j$ is a positive set. Minimum such $t$ gives us a positive minimal set.
This completes the proof.
\qed

\begin{lemma}\label{correctness}
Suppose that $I^{+}$ is a positive subset with $bg(I^+) > K$
and $I^{-}$ is a negative subset where $bg(I^{-}) \le K$ and $I^+ \cap I^- \ne\emptyset$.
If $bg( I^{+} \cup I^{-}) \le K$ then $I^{+} \cup I^{-}$ forms a positive subset.
\end{lemma}
\pf Let $X=I^-\cap I^+$. By the assumption that $I^+$ can be processed after processing
$I^-$ we have $bg(I^+\setminus X)\leq K-|I^-|+|N^*(I^-)|$. On the other hand, since $bg(I^+)>K$, we have
$K-|X|+|N^*(X)|<bg(I^+\setminus X)$. From these two we conclude that:\\
\begin{equation}\label{eq}
|N^*(I^-)|>|I^-|-|X|
\end{equation}

Moreover, because $I^+$ is a positive set then $|N^*(I^+)|\geq |I^+|$.
By (\ref{eq}), $I^+$ being positive, and the fact that
$|N^*(S \cup T)| \geq |N^*(S)| + |N^*(T)|$ for any $S$ and $T$, we have
$|N^*(I^+\cup I^-)| \geq |N^*(I^+)|+|N^*(I^-)|\geq |I^+|+|I^-|-|X| = |I^+\cup I^-|$, i.e., $I^+\cup I^-$ is a positive subset.
\qed \\

Given a bipartite graph $H$, Lemma \ref{lm:minpos} suggests a basic strategy,
if there are positive sets, find a positive minimal subset $I$, process it.  When a given subset $I$ is processed, we
would consider the remaining bipartite graph and again try to find a positive
minimal subset to process, if one exists. Note that
$H\setminus (I\cup N^*(I))$ may have positive sets even if $H$ does not.
For example, in the graph of Figure \ref{fig:2},
$H'=(J_2\cup J_1 \cup J, D \cup E \cup F)$ has no positive set,
but $J$ is positive in $H'\setminus J_1$.  When a
subset $I\subseteq B$ is processed we generally would like to process
any sets that are positive in the remaining bipartite graph.  That is,
we would like to process $c\ell(I)$, defined below.
For our purpose we order all the prime sets lexicographically, by assuming some ordering on the vertices of $B$.

\begin{dfn}\label{def:closure}
  Given current budget $K$ and given $I\subseteq B$, let $c\ell_K(I)=\cup_{i=1}^{r} I_i \cup I $ where each $I_i \subseteq B$,
  $1 \le i \le r$ is the lexicographically first
   positive minimal subset in
   $H_i=H \setminus (\cup_{j=0}^{i-1} I_j \cup N^*(\cup_{j=0}^{i-1} I_j)) $
   ($I_0=I$) such that in $H_i$ we have $bg(I_{i}) \le K-|\cup_{j=0}^{i-1} I_j|+|N^*(\cup_{j=0}^{i-1} I_j)|$.  Here $r$ is the number of times the process of processing a positive minimal
set can be repeated after processing $I$.
\end{dfn}
When the initial budget $K$ is clear from context, we use $c\ell(I)$ rather than $c\ell_K(I)$.
Note that $c\ell(I)$ could be only $I$, in this case $r=0$. For instance consider Figure \ref{fig:2}. In the graph induced by $\{J,J_1,J_2,I,D,E,F,L,P\}$ we have $c\ell(J)=J\cup J_1$ with respect to any current budget $K$ at least $12$.
%Note that $c\ell(I)$ is well-defined.

\section{Polynomial Time Algorithm for Trivially Perfect Bipartite and Co-bipartite Graphs}\label{sec:poly-TP-CB}

In this section we define trivially perfect bipartite graphs and co-bipartite graphs,
and discuss the key properties that are used in our algorithm for solving
the bipartite graph ordering problem in these bipartite graphs.  In particular, it is
possible to enumerate the prime sets of these graphs by looking at a way
to construct the graphs with a tree of graph join and union operations.

%Arash
The subclass of trivially perfect bipartite graphs called {\em laminar family bipartite graphs} were considered in \cite{MSW10} to obtain a
polynomial time approximation
scheme (PTAS) for special instances of a job scheduling problem. Each instance of the problem in \cite{MSW10} is a bipartite graph
$H=(J,M,E)$ where $J$ is a set of jobs and $M$ is a set of machines. For every pair of jobs $i,j \in J$ the set of machines that can process
$i,j$ are either disjoint or one is a subset of the other. The trivially perfect bipartite graphs also play an important role in
studying the list homomorphism problem. The authors of \cite{EKLT12} showed that for these
bipartite graphs, the list homomorphism problem can be
solved in logarithmic space. They were also considered in the fixed parametrized version of the list homomorphism problem
in \cite{CEM13}.

We call these bipartite graphs ``trivially perfect
bipartite graphs'' because the definition mirrors one of the equivalent
definitions for trivially perfect graphs.

\begin{dfn}[trivially perfect bipartite graph, co-bipartite graph]
  \label{def:tp}
  A bipartite graph $H=(B,S,E)$ is called \emph{trivially perfect} , respectively
  a \emph{co-bipartite graph} if it can be constructed by applying the
  following operations.
  \begin{itemize}

  \item A bipartite graph with one vertex is both trivially perfect and a co-bipartite graph.
  \item If $H_1$ and $H_2$ are trivially perfect then the disjoint union of $H_1$ and $H_2$ is trivially perfect.

    Similarly, the disjoint union of co-bipartite graphs is also a co-bipartite graph.
  \item If $H_1$ and $H_2$ are trivially perfect then by joining
  %\footnote{
  %  \cite{EKLT12} used the term ``special sum'' for this operation which we
  %  call a ``join''.}
  every sold vertex in $H_1$ to every bought vertex in $H_2$, the resulting bipartite graph is trivially
    perfect.

    If $H_1$ and $H_2$ are co-bipartite graphs, their \emph{complete} join---where every sold
    vertex in $H_1$ is joined to every bought vertex in $H_2$ and every bought vertex in $H_1$ is
    joined to every sold vertex in $H_2$---is a
    co-bipartite graph.
  \end{itemize}
\end{dfn}

\begin{figure}[htbp]
 \begin{center}
\includegraphics[scale=0.4]{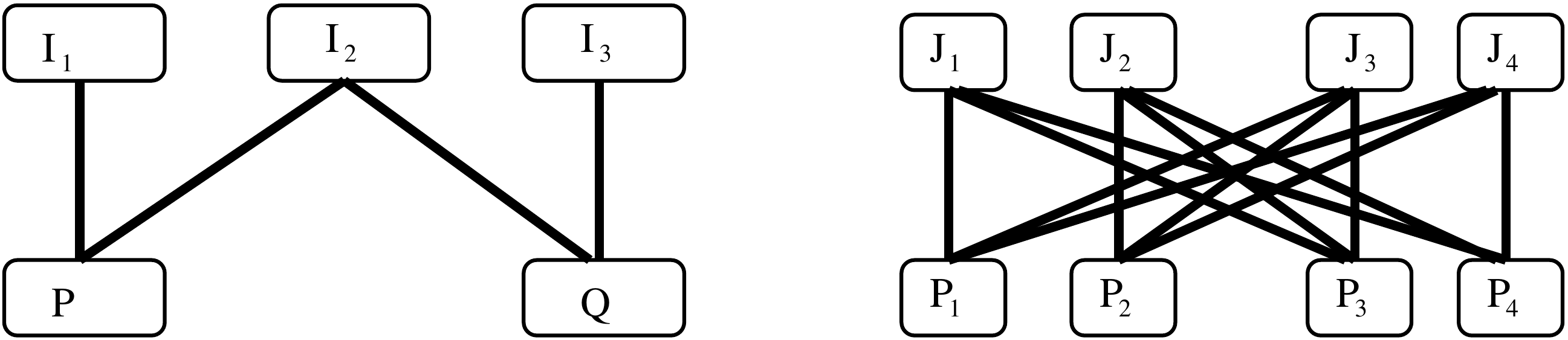}
\caption{
Each bold line shows a complete connection, i.e. the induced sub-graph
by $I_1 \cup P$ is a biclique.
}
    \label{fig:sibling}
  \end{center}
 \end{figure}

An example of each type of graph is given in Figure \ref{fig:sibling}. In the left figure (trivially perfect) $I=\{I_1,I_2\}$ and $J=\{I_2,I_3\}$ are prime sets.
On the right figure (co-bipartite graph) prime sets are $R_1=\{J_1,J_2,J_3\}, R_2=\{J_1,J_2,J_4\},R_3=\{J_3,J_4,J_1\},R_4=\{J_3,J_4,J_2\}$ are prime
sets.% and for example $R_1,R_2$ are sibling.

These two classes of bipartite graphs can be classified by forbidden obstructions, as follows.

\begin{lemma}[\cite{EKLT12,GV97}] \label{lm:obstruct}
  $H$ is trivially perfect if and only if it does not contain any of the
  following as an induced sub-graph: $C_6$, $P_6$.

  $H$ is a co-bipartite graph if and only if it does not have any of the
  followings as an induced sub-graph %$P_7$, $Star-(1,2,3)$ with vertex set $V=\{1,2,3,4,5,6,7\}$ and
  %$E=\{\{1,2\},\{2,3\},\{3,4\},\{4,5\} \{5,6\},\\ \{3,7\}\}$,  Sun$(4)$ with vertex set $V$ and
  %$E=\{\{1,2\},\{2,3\},\{3,4\},\{4,1\},\{1,5\},\{2,6\},\{3,7),\{4,8\}\}$.
  %(see Figure \ref{fig:co-bipartite graph}).

\begin{figure}[htbp]
 \begin{center}
\includegraphics[scale=0.7]{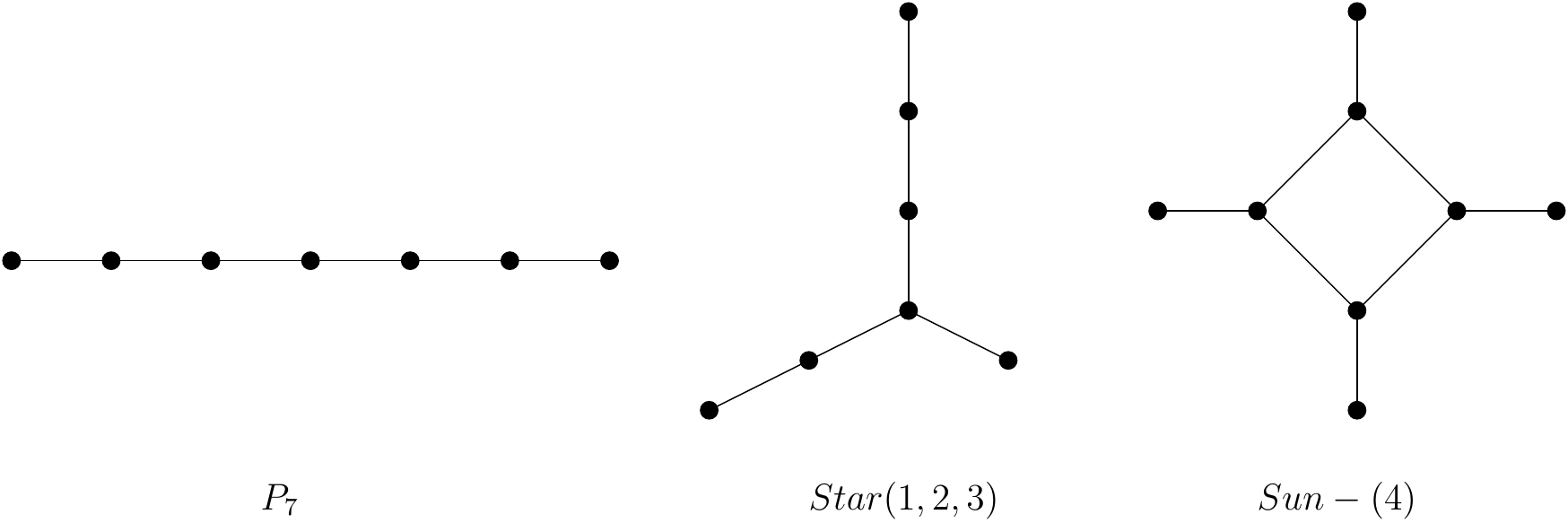}
\caption{
\footnotesize{Forbidden constructions for co-bipartite graphs.
 }
}
  \end{center}
 \end{figure}
\end{lemma}
Our algorithm to solve $bg(H)$ for trivially perfect bipartite graphs and co-bipartite graphs
centers around constructing $H$ as in Definition \ref{def:tp}.  We view this
construction as a tree of operations that are performed to build up the final
bipartite graph, and where the leaves of the tree of operations are \emph{bicliques}.
If $H$ is not connected then the root operation in the tree is a disjoint
union, and each of its connected components is a trivially perfect
bipartite graph (respectively co-bipartite graph).  If $H$ is connected, then the root operation is a
join. The following lemma shows how to find such a decomposition tree for given trivially perfect bipartite graph in polynomial time. For co-bipartite graph $H$ a polynomial time algorithm to compute decomposition tree is given in \cite{GV97}.

\begin{lemma}\label{lm:make-tree}
  Given a trivially perfect bipartite graph $H$ with $n$ vertices, there exists an algorithm that finds a decomposition tree for $H$ in 
  time $O(n^3)$.
\end{lemma}
\pf %First suppose $H=(B,S,E)$ is a trivially perfect bipartite graph with $n$ vertices.
If $H$ is not connected then the root of tree $T$ is $H$ and
two children $H_1,H_2$ are chosen such that $H_1$ contains all the connected components $H'=(B',S')$ of $H$ where $|B'| < |S'|$ (if there is any)
and $H_2$ contains all the other connected components. The root has a label "union". Note that if there exists only one such $H'$ then $H_1=H'$.
If for every connected component of $H$ the size of its bought vertices is smaller than the size of its sold vertices then one of them would be in $H_2$ and the rest lie in $H_1$.

If $H$ is connected then we proceed as follows.
Let $1 \le m \le n$ be a maximum integer such that the following test passes.
Let $B_2$ be the set of vertices in $B$ which have degree at least $m$ and let $B_1=B \setminus B_2$. Let $S_1$ be the set of all vertices in $S$
that are common-neighborhood of all the vertices in $B_2$. If $|S_1| < m$ then the test fails. Moreover, if there exists a vertex $v \in B_1$ such that $N(v) \not\subset S_1$ then
the tests fails. %If there exists a vertex in $B_1$ with degree at least $m$ then the test fails.
If the test passes then let $S_2=S \setminus S_1$ and let the root of $T$ be $H$ with label "join" and the
left child of $H$ is $H[B_1 \cup N(B_1)]$ and the
right child of the root is $H_2= H \setminus H_1$. Note that by the definition of trivially perfect bipartite graphs.
If the test fails for every $m$ then $H$ is not trivially perfect.

We continue the same procedure from each node of the tree until  each node becomes a biclique. Node that $T$ has at most $n$ nodes. For a particular $m$, checking all the conditions of the test takes
$O(n)$. Therefore the whole procedure takes $O(n^3)$.
 \qed \\

\begin{figure}[htbp]
 \begin{center}
\includegraphics[scale=0.7]{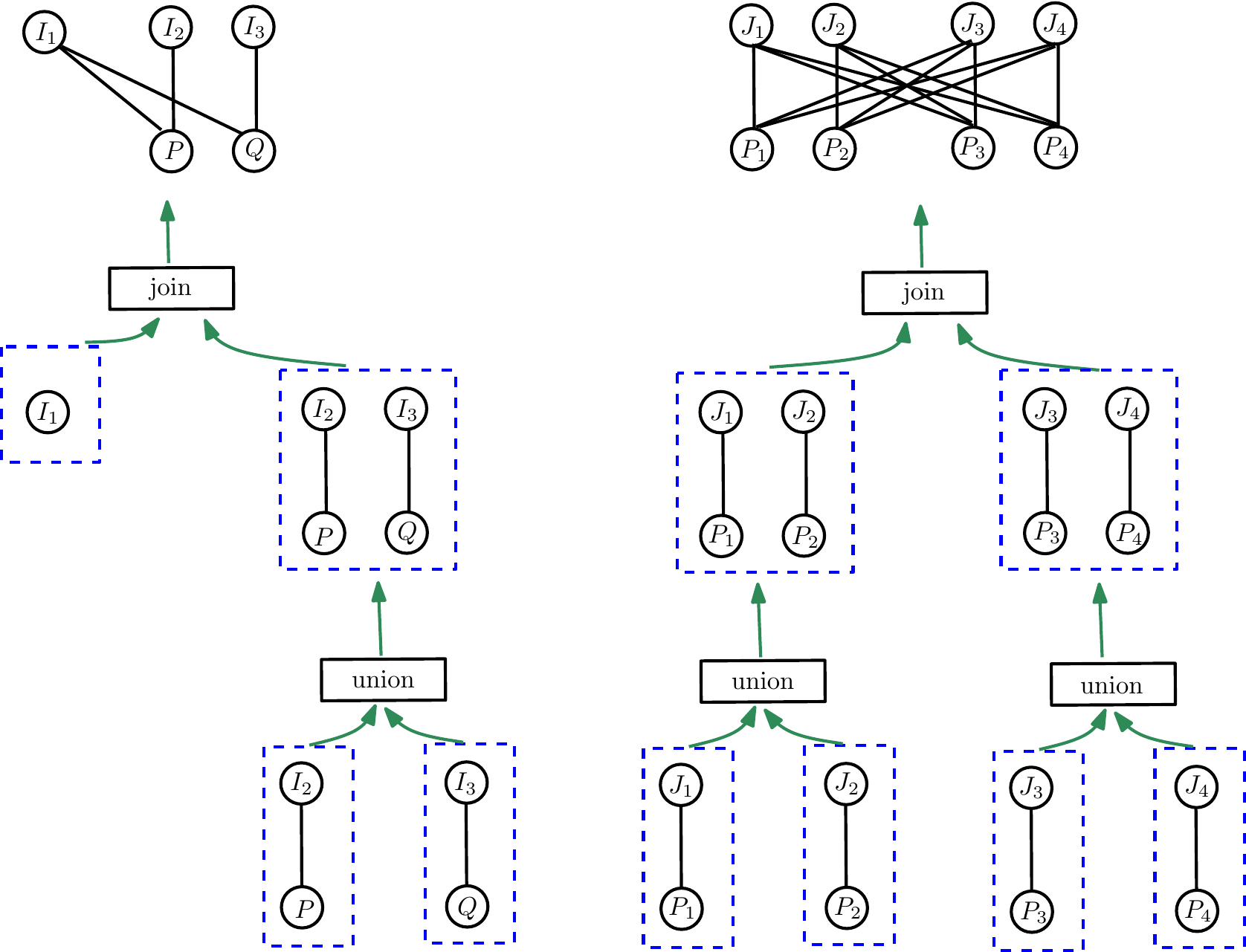}
\caption{
Decomposition tree associated to the graphs in Figure \ref{fig:sibling}.
}
    \label{fig:decomposition}
  \end{center}
 \end{figure}

%%%%%%%%%%%%%%%%%%%%%%%%%%%%%%%%%%%%%%%%%%%%%%%%%%%%%%%%%%%%%%%%
Algorithm \ref{budget-poly} shows that how we traverse a decomposition tree in bottom-up manner and for each node of the tree we do a binary search to find the optimal budget for the graph associated to that node. Note that we assume for the graph associated to a particular node of tree the optimal budgets for its children have been computed and stored.
\begin{algorithm}[H]
\begin{algorithmic}[1]
\State \textbf{Input:} Trivially perfect bipartite (resp.) graph $H=(B,S,E)$, its decomposition tree $T$\;
\State \textbf{Output:} $bg(H)$\;
\State Start from leaves of $T$ and traverse $T$ in bottom-up manner:\;
\State Let $H_x=(B_x,S_x,E_x)$ be the associated graph to node $x$ of $T$\;
\State\Comment{Assume optimal strategies for children of $H_x$ are already computed}
\State $l=1$ and $h=|B_x|$
\While{ $l\leq h$ }
	\If{\textsc{BudgetTriviallyPerfect}$(H_x,\lfloor\frac{l+h}{2}\rfloor)$ (resp. BudgetCo-Bipartite) is True}
    	\State $h=\lfloor\frac{l+h}{2}\rfloor$ \;
        \Else
        	\State $l=\lfloor\frac{l+h}{2}\rfloor+1$
    \EndIf
\EndWhile
\Return l\;
\end{algorithmic}
\caption{{\textsc{BudgetTPC}} ( $H, K$)}
\label{budget-poly}
\end{algorithm}

If the graph is constructed by union operation it requires a merging function. Such a function is given
in Algorithm \ref{combine}. \textsc{Combine} function takes optimal solutions of two trivially perfect (respectively co-bipartite)
graphs and return an optimal strategy for the union of them. In what follows, we give the  description of our algorithm and prove its correctness.

Recall that we assume eavery vertex in $B$ has at least one neighbor. 

% \begin{remark}\label{how-the-tree}
% If $H$ is not connected then the root of tree $T$ is $H$ and
% two children $H_1,H_2$ are chosen such that  $H_1$ contains all the connected components
% $H'=(B',S')$ of $H$ where $|B'| < |S'|$ (if there is any) and $H_2$ contains all the other connected components. 
% \end{remark}

%%%%%%%%%%%%%%%%%%%%%%%%%%%%%%%%%%%%%%%%%%%%%%%%%%%%%%%%%%%%%
%%%%%%%%%%%%%%%%%%%%%%%%%%%%%%%%%%%%%%%%%%%%%%%%%%%%

\begin{algorithm}[H]
\begin{algorithmic}[1]
\State \textbf{Input:} Trivially perfect bipartite graph $H=(B,S,E)$ and budget $K$\;

\Comment We assume decomposition tree $T$ of $H$ is given.

%$H_1=(B_1,S_1,E_1)$ and $H_2=(B_2,S_2,E_2)$, $bg(H_1)$, $bg(H_2)$, its decomposition tree $T$, and budget $K$\;
\State \textbf{Output:} "True" if we can process $H$ with budget at most $K$, otherwise "False".\;
\If{$S= \emptyset$ and $K \ge 0$ } \Return True\;
\EndIf

\If{$H$ is a bipartite clique and $|B|\leq K$ } 
process $H$ by ordering vertices in $B$ first and then ordering vertices in $S$ after and \Return True\;
\EndIf
%
% \If{ $|I| > K$ for all prime $I \subseteq B$}
% \Return False\;
% \EndIf
% \If {there is a positive minimal subset $I$ with $bg(I) \le K$} %\label{Alg-general-thirdIF}
%      \Return Process $I$ and $N^*(I)$, call \textsc{BudgetTriviallyPerfect} $(H[B \setminus I,S \setminus N^{*}(I)],K-\vert I\vert +\vert N^{*}(I)\vert)$\;
% \EndIf %\label{Alg-gnr-end-thirdif}
% \If{a positive set $I$ with the smallest budget has $bg(I) > K$ }
% \Return False\;
% \EndIf \label{lastIF}
\If {$H$ is constructed by join operation between $H_1=(B_1,S_1)$ and $H_2=(B_2,S_2)$}\label{if-join}

\Comment $bg(H_1),bg(H_2)$ already computed and assume $B_1$ and $S_2$ induce a bipartite clique.

\If { $bg(H_1) > K$} \Return False;
\ElsIf {$bg(H_2) > K-|B_1|+|S_1|$} \Return False;
\Else {}  first process $H_1$ then process $H_2$ and \Return True,
\EndIf
\EndIf \label{if-join-end}

\If{$H$ is constructed by union of $H_1$ and $H_2$}

 \If {$\exists$ a positive minimal subset $I$ with $bg(I) \le K$} %\label{Alg-general-thirdIF}
     \State Process ($I,N^*(I)$),
     \State \Return call \textsc{BudgetTriviallyPerfect}
     $(H[B \setminus I,S \setminus N^{*}(I)],K-\vert I\vert +\vert N^{*}(I)\vert)$\;
\EndIf %\label{Alg-gnr-end-thirdif}
\If{a positive set $I$ with the smallest budget has $bg(I) > K$ }
\Return False\;
\EndIf \label{lastIF}

%    \If { $H_1$ is postitive and $bg(H_1) < K$ }
%
%  %\State Process $H_2$
%
%     \If {$ bg(H_2 ) > K-|B_1|+|S_1|$} \Return False;
%
%     \Else { } Process $H_1$ and then process $H_2$ and \Return True;
%
%     \EndIf
%
%   \EndIf
%
%   \If { $H_1$ is positive and $bg(H_1) > K$ }
%
%     \If { $H_2$ is negative } \Return False;
%
%
%
%  %\EndIf
%
%     \Else { } \While{ $\exists$ a positive component $H'=(B',S')$ of $H$ with $bg(H') \le K$}
%
% 	      \State Process $H'$ and set $K=K-|B'|+|S'|$ and set $H \leftarrow H \setminus H'$.
%
% 	      \EndWhile
%
% 	      \If { $H \ne \emptyset$ } \Return False;
%               \Else { } \Return True;
%
% 	      \EndIf
%     \EndIf
%
%   \EndIf

 %\ElsIf {there is a connected $H'$ of $H_1 \cup H_2$ which is positive and $bg(H') < K$}

 %  \State Process $H'$ and Call \textsc{BudgetTriviallyPerfect} ($H \setminus H',K-|B_2|+|S_2|$)

 \If{ $bg(H_1)>K$ OR $bg(H_2)>K$ }
  \Return False\;

  %\Comment $H_1,H_2$ both negative

 \Else {} \Return \textsc{Combine}($H_1,H_2, K$)
 \EndIf

\EndIf

\end{algorithmic}
\caption{{\textsc{BudgetTriviallyPerfect}} ($H, K$)}
\label{alg:trivially-perfect}
\end{algorithm}

\begin{algorithm}[H]
\begin{algorithmic}[1]
\State \textbf{Input:} Optimal strategies for $H_1=(B_1,S_1,E_1), H_2=(B_2,S_2,E_2)$ and budget $K$\;
\State \textbf{Output:} "True" if we can process $H=H_1\cup H_2$ with budget at most $K$, otherwise "False".\;

\State Let $J_1$ be the first prime set in $H_1$ and $J_2$ be the first prime set in $H_2$.

\If {$|J_1| > K$ OR  $bg( H_2) > K- |c\ell(J_1)|+|N^*(c\ell(J_1))| $}
	\State Process $c\ell(J_2)$ and $N^*(c\ell(J_2))$
       \State Call $\textsc{Combine} (H_1, H_2 \setminus (c\ell(J_2) \cup N^*(c\ell(J_2))), K-|c\ell(J_2)|+|N^*	(c\ell(J_2))| ) $.

\Else{}  \State Process $ c\ell(J_1)$ and $N^*(c\ell(J_1))$,
         \State Call \textsc{Combine} $(H_1\setminus (c\ell(J_1) \cup N^*(c\ell(J_1))), H_2, K-|c\ell(J_1)|+|N^*(c\ell(J_1))| ) $.

\EndIf

\end{algorithmic}
\caption{{\textsc{Combine}} ( $H_1,H_2, K$)}
\label{combine}
\end{algorithm}

\begin{tm}\label{thm:poly-TP-CB}
  For trivially perfect bipartite graphs $H$ with $n$ vertices the \textsc{BudgetTriviallyPerfect} algorithm runs in
  $O(n^2)$ and correctly decides if $H$ can be processed with budget $K$ (Algorithm \ref{alg:trivially-perfect}).
\end{tm}
\pf The correctness of lines 3,4 is obvious. It is clear that if $H$ is obtained from $H_1$ and $H_2$ by join operation then 
the any optimal strategy must starts with $H_1$.
Therefore the Lines 5--8 are correct.

Suppose $H$ is obtained from $H_1,H_2$ by union operation. Let $I$ be a positive minimal set and let $H'$ be the induced sub-graph
of $H$ by $I \cup N^*(I)$. If $H'$ is not connected then there is at least one connected component of $H'$
that is positive, a contradiction to minimality of $I$. Thus we may assume $H'$ is connected. According to the decomposition of $H'$ there are
$H'_1=(B'_1,S'_1)$ and $H'_2=(B'_2,S'_2)$ such that $H'_1$ and $H'_2$ are trivially perfect bipartite graph.
% (respectively co-bipartite graphs)
Suppose every bought vertex in $B'_1$ is adjacent to every sold vertex in $S'_2$.
 Observe that  any positive set must include either a positive part of $H'_1$ or all $H'_1$ together with a positive part of $H'_2$.
In the former case, we search in $H'_1$ for a positive set. In the later one, we search for a positive set $I'$ in $H'_2$
so that $|N^*(I')|-|I'|\geq |B'_1|-|S'_1|$. In either case, we repeat the same procedure and
traverse the decomposition tree to find a positive set. This takes $O(n^2)$. The correctness of Lines 10-12 is followed by Lemma 
\ref{first-prime} and \ref{lm:minpos}. 
Suppose line 13 is incorrect and all positive subsets would have budget above $K$.
Let $I^+$ be one such subset. Then there would be a
way to process $I^+$ with budget at most $K$ in $H$. In that case, we would process some negative set
$I^-$ which somehow reduces the budget of processing $I^+$; this can
only be so if $I^-\cap I^+ \neq \emptyset$. In this case the Lemma
\ref{correctness} states that $I^+ \cup I^-$
is itself a positive set with budget at most $K$, a contradiction.\\

%Note that by Lemma \ref{positive-TPC}, in polynomial time we can find positive minimal sets.

We continue our argument by assuming that $H$ is constructed form $H_1=(B_1,S_1)$ and $H_2=(B_2,S_2)$ by "union" operation.
%In this case, we call \textsc{Combine} function.
We proceed by showing the correctness of \textsc{Combine} function. Let $J_1$ be the first
prime set in $H_1$ and $J_2$ be the first prime set in $H_2$. 

%The following
%claim shows that there is an optimal ordering for $H=H_1\cup H_2$ such that either $J_1$ or $J_2$ is the first prime to process. %According to Steps \ref{if-combine} and \ref{else-combine} in \textsc{Combine} function, there is an optimal strategy for $H$ that starts by processing $J_1$ or $J_2$.

\begin{obs}
 Let $H_1=(B_1,S_1)$ and $H_2=(B_2,S_2)$ be two disjoint trivially perfect bipartite graphs ($H_1\cap H_2=\emptyset$). Suppose optimal
strategies for computing the budget for $H_1$ and $H_2$ are provided.  If $J_1, J_2$ are the first prime sets in $H_1,H_2$ then there is an optimal
ordering for $H=H_1\cup H_2$ such that either $c\ell(J_1)$ or $c\ell(J_2)$ is processed first. 
\end{obs}

% 
% 
% \begin{claim}\label{union-first-prime}
% Let $H_1=(B_1,S_1)$ and $H_2=(B_2,S_2)$ be two disjoint trivially perfect bipartite graphs ($H_1\cap H_2=\emptyset$). Suppose optimal
% strategies for computing the budget for $H_1$ and $H_2$ are provided.  If $J_1, J_2$ are the first prime sets in $H_1,H_2$ then there is an optimal
% ordering for $H=H_1\cup H_2$ such that either $J_1$ or $J_2$ is the first prime.
% \end{claim}
% \pf Without loss of generality, we assume $H_1$ and $H_2$ are connected. The claim is correct for the case when $H_1$ and $H_2$ are bipartite cliques.
% Our proof is based on induction. By definition, let $H_i$
% be constructed from $H_{i1}$ and $H_{i2}$ by join operation where every sold vertex in $H_{i2}$ is
% adjacent to every bought vertex in $H_{i1}$, for $i=1,2$.
% Accordingly, $J_1\in H_{11}$ and $J_2\in H_{21}$. It implies that
% we can reduce the problem to finding the first prime set to process
% either in $H_{11}$ or in $H_{21}$ that are trivially perfect bipartite graphs.
% By induction hypothesis, for $H_{11}\cup H_{12}$ the first prime is either the first prime in $H_{11}$ or
% the first prime in $H_{12}$. It completes the proof.
% \qed \\

%We show that there exists
To complete the proof for correctness of \textsc{Combine} function, it remains to show that the Combine function correctly chooses
between $J_1$ and $J_2$, the first prime set to process in $H$. Suppose we have  $|J_1| < K$ and $bg( H_2) \le K- |c\ell(J_1)|+|N^*(c\ell(J_1))|$.
We claim there exists an optimal strategy for $H$ that starts processing $c\ell(J_1)$ first. Let $opt$ be the optimal strategy that process $c\ell(J_2)$ first.
Let $J_2,J_3,\dots,J_k$ be the prime subsets in $H_2$ that are processed by $opt$ before starting $J_1$ in $H_1$ (note that by Observation above
$opt$ starts processing $c\ell(J_1)$ in $H_1$ first). We note that $bg(H_2) \ge bg(J_2 \cup J_3 \cup \dots \cup J_k)$. Because we assume that there is no positive set
in $H_2$. Therefore we have
$K-|c\ell(J_1)|+|N^*(c\ell(J_1))| \ge bg(H_2) \ge bg(J_2 \cup J_3 \cup \dots \cup J_k)$ and hence we obtain a strategy $opt'$ that starts
with $c\ell(J_1)$ first and then it processes $J_2,J_3,\dots,J_k$ from $H_2$ and then it follows $opt$. Observe that under $opt'$, $bg(H)$ does not increases.

Note that finding $c\ell(J)$ takes $O(n)$ and it can be determined according to join or union operation as follows.  
 
%Note, we still need to show how to keep track of $c\ell(J)$ for different prime sets $J$ and answer queries such as $bg( c\ell(J)) \le K$.
%Since all prime sets $J$ must be leaves
%of the decomposition tree, there are only a polynomial number of prime sets to consider
%and by Lemma \ref{lm:make-tree} we can determine what they are. It takes $O(n)$ time to traverse the decomposition tree. Now,
%it requires to update $c\ell(J)$ as we traverse the decomposition tree in bottom-up manner. 
Suppose $H$ is associated to a
node of the decomposition tree and it is constructed from $H_1$ and $H_2$ either by union or join operation.
Without loss of generality, we assume there is no positive minimal set in $H$, as otherwise, we process them first.
Let $c\ell(J) \subseteq B_1$. First, if the operation is union then $c\ell(J)$ does not change. Second, suppose the operation is join and every sold vertex in
$H_2$ is adjacent to every bought vertex in $H_1$. If $c\ell(J)$ is the entire $B_1$ then $c\ell(J)$ is $B_1$ plus all positive minimal sets in $H_2$.
If $c\ell(J)\subset B_1$ then it does not change in $H$.
Therefore, updating $c\ell(J)$ at each step takes at most $O(n)$ time. These would imply that the overall running time would be $O(n^2)$. 
%We also observe that we can keep track of $bg(H')$ where $H'$ is any of the bipartite
%graphs in the decomposition tree.
\qed \\

In what follows we show that there is a subclass of trivially perfect bipartite graphs that are also circle bipartite graphs.
A bipartite graph $H=(B,S,E)$ is called a \textit{chain graph} if the neighborhoods of vertices in $B$ form a chain, i.e, if there is an ordering of vertices in $B$, say $w_1,w_2,\cdots,w_p$, such that $N(w_1)\supseteq N(w_2)\supseteq \cdots \supseteq N(w_p)$.

\begin{figure}[htbp]
\begin{center}
\includegraphics[scale=0.7]{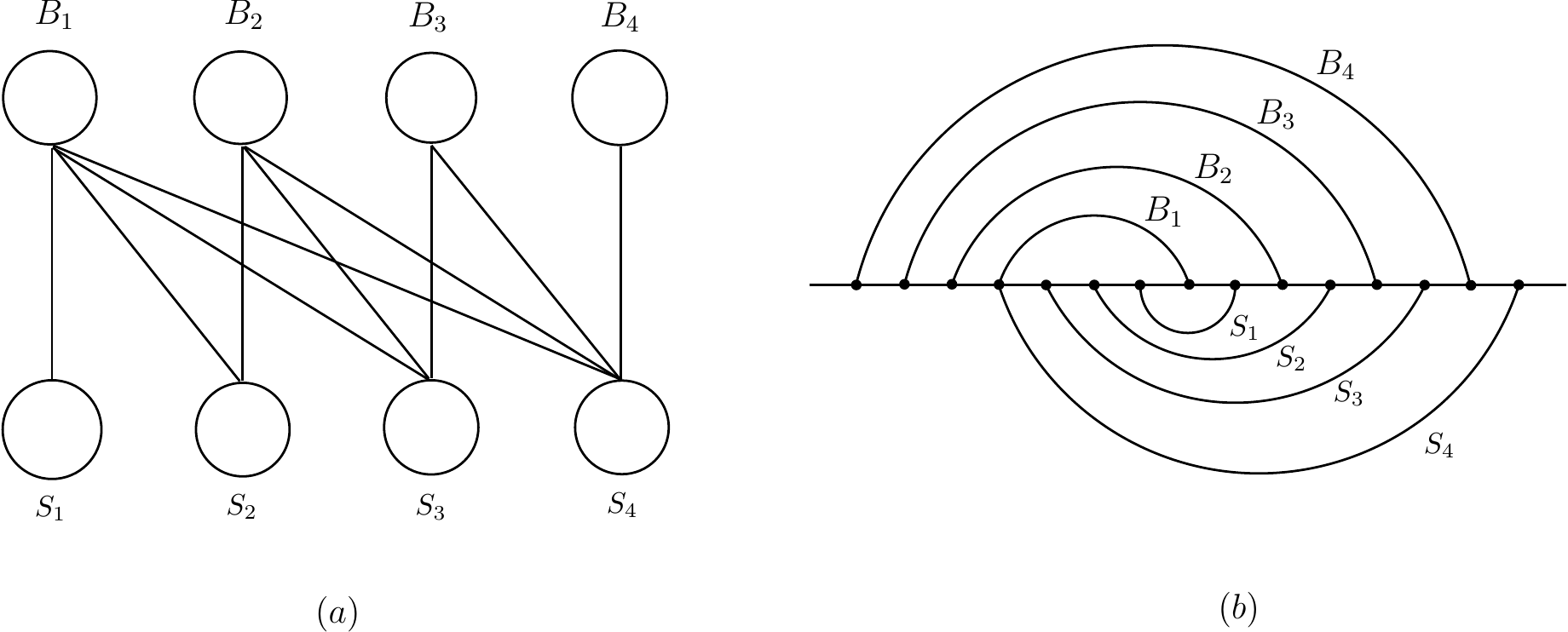}
\caption{
$(a)$: Each bag $B_i$ and $S_i$ contains at least one vertex, for $1\leq i\leq 4$.
A line between $B_i$ and $S_j$ means that vertices in $B_i\cup S_j$ induce a complete bipartite graph for $1\leq i,j\leq 4$.
$(b)$ :
Each $B_i$ and $S_i$ arc represents a collection of arcs; the number of arcs which are represented by each $B_i$ and $S_j$ arc is equal to the number of
vertices in bag $B_i$ and $S_i$, for $1\leq i, j\leq 4$.
}
\label{fig:4}
\end{center}
\end{figure}

It is easy to see that the neighborhoods of
vertices in $S$ also form a chain. \textit{Chain graphs} are subsets of
both \textit{trivially perfect bipartite graphs} and \textit{circle bipartite graphs}.
Any \textit{chain graph} can be visualized as what is
depicted in Figure \ref{fig:4}(a), and the corresponding RNA model for the bipartite graph ordering problem looks like Figure \ref{fig:4}(b).

Next we present a polynomial time algorithm for co-bipartite graphs. Our algorithm for this class of graphs is quite similar to Algorithm \ref{alg:trivially-perfect}. The main difference is in the way we deal with co-bipartite graph $H=(B,S,E)$ when it is constructed from two co-bipartite graphs $H_1=(B_1,S_1,E_1)$ and $H_2=(B_2,S_2,E_2)$ by join operation. Recall that in join operation for co-bipartite graphs, $H[B_1\cup S_2]$ and $H[B_2\cup S_1]$ are bipartite cliques. Observe that in this case there are two possibilities for processing $H$:
\begin{itemize}
\item first process entire $B_2$ then solve the problem for $H_1$ with budget $K-|B_2|$, and at the end process $S_2$, or
\item first process entire $B_1$ then solve the problem for $H_2$ with budget $K-|B_1|$, and at the end process $S_1$.
\end{itemize}

For the case when $H$ is constructed from $H_1$ and $H_2$ by union operation we call \textsc{Combine} function (Algorithm \ref{combine}). The description of our algorithm is given in Algorithm \ref{alg:co-bipartite}. The proof of correctness of Algorithm \ref{alg:co-bipartite} is almost identical to the proof of Theorem \ref{thm:poly-TP-CB}.

\begin{tm}
Algorithm \ref{alg:co-bipartite} in polynomial times decides if co-bipartite graph $H$ can be processed with budget at most $K$.
\end{tm}

\begin{algorithm}[H]
\begin{algorithmic}[1]
\State \textbf{Input:}  Co-bipartite graph $H=(B,S,E)$ constructed from
$H_1=(B_1,S_1,E_1)$ and $H_2=(B_2,S_2,E_2)$, $bg(H_1)$, $bg(H_2)$, its decomposition tree $T$, and budget $K$\;
\State \textbf{Output:} "True" if we can process $H$ with budget at most $K$, otherwise "False".\;
\If{$S= \emptyset$ and $K \ge 0$ OR $H$ is a bipartite clique and $|B|\leq K$} \label{firstIF-co}
 process $H$ and \Return True\;
\EndIf
% \If{ $|I| > K$ for all prime $I \subseteq B$}
% \Return False\;
% \EndIf
% \If {there is a positive minimal subset $I$ with $bg(I) \le K$}
%      \Return Process $I$ and $N^*(I)$, call \textsc{BudgetCo-Bipartite} $(H[B \setminus I,S \setminus N^{*}(I)],K-\vert I\vert +\vert N^{*}(I)\vert)$\;
% \EndIf
% \If{a positive set $I$ with the smallest budget has $bg(I) > K$ }
% \Return False\;
% \EndIf \label{lastIF-co}
%\State Let $H$ be constructed from $H_1=(B_1,S_1,E_1)$ and $H_2=(B_2,S_2,E_2)$ according to decomposition tree $T$\;
\If {$H$ is constructed by join operation between $H_1$ and $H_2$}
	\If{ $bg(H_1) \leq K-|B_2|$ }
    	\Return True and process $B_2$, process $H_1$ with
    	
    	\hspace{3mm}  budget $K-|B_2|$, and process $S_2$\;
    	
    \ElsIf { $bg(H_2)\leq K-|B_1|$ }
    	\Return True and process $B_1$, process $H_2$ with
    	
    	\hspace{3mm} budget $K-|B_1|$, process $S_1$\;
    \Else {} \Return False\;
    \EndIf
\EndIf 

\If{$H$ is constructed by union of $H_1$ and $H_2$}

 \If {$\exists$ a positive minimal subset $I$ with $bg(I) \le K$} %\label{Alg-general-thirdIF}
     \State Process $I$ and $N^*(I)$,
     \State \Return call \textsc{BudgetCo-Bipartite}
     $(H[B \setminus I,S \setminus N^{*}(I)],K-\vert I\vert +\vert N^{*}(I)\vert)$\;
\EndIf %\label{Alg-gnr-end-thirdif}
\If{a positive set $I$ with the smallest budget has $bg(I) > K$ }
\Return False\;
\EndIf

\If{ $bg(H_1) >K$ OR $bg(H_2)>K$}
\Return False\;
%\EndIf
\Else {} \Return \textsc{Combine}( $H_1,H_2, K$)
\EndIf
\EndIf
\end{algorithmic}
\caption{{\textsc{BudgetCo-Bipartite}} ($H, K$)}
\label{alg:co-bipartite}
\end{algorithm}

\section{Polynomial Time Algorithm for Bipartite Permutation Graphs}\label{sec:poly-permutation}
A bipartite graph $H=(B,S,E)$ is called permutation graph (proper interval bipartite graph) if there exists an ordering $<$
of the vertices in $B$ such that the neighborhood of each vertex in $S$ consists of consecutive vertices in $<$. Moreover, for any two
vertices $s_1,s_2 \in S$ if $N(s_1) \subset N(s_2)$ then the last neighbor of $s_1$ and the last neighbor of $s_2$ are the same.
These bipartite graphs were exactly those bipartite graph for which  the  minimum cost homomorphism problem can be solvaled in polynomial time \cite{GHRY08}. 
They are also studied  in job scheduling problems \cite{KKRS13,MS12}.

We refer to a set of consecutive vertices in such an ordering as an \textit{interval}. Figure
\ref{fig:2} is an example of a bipartite permutation
graph.

Note that the class of circle bipartite graphs $G=(X,Y)$, for which
obtaining the optimal budget is NP-complete, contains the class of
bipartite permutation graphs.

We obtain an ordering $<$ for vertices in $S$ by setting $s < s'$ if the first neighbor of
$s$ is before the first neighbor of $s'$ in $<$ as therwise $s' < s$. Let $b_1,b_2,\dots,b_p$ and $s_1,s_2,\dots,s_q$ be the orderings of $B$ and $S$
respectively. If $s_ib_j$ and $s_{i'}b_{j'}$ are edges of $H$ and $j' <j$ and $i < i'$ then $s_ib_{j'},s_{i'}b_j \in E(H)$. 
Such an ordering is called {\em min-max} ordering \cite{GHRY08}.

 Let $B[i,j]$ denote the interval of vertices 
$b_i,b_{i+1},\dots,b_j$. In the Algorithm \ref{alg:permutation} we compute the optimal budget for every $B[i,j]$. In order to compute $bg(B[i,j])$ we assume that the optimal 
strategy starts with some sub-interval $J$ of 
$B[i,j]$ and it processes $c\ell(J)$. Then we are left with two disjoint instances $B_1,B_2$ (this is because of property of the min-max ordering). We then argue
how to combine the optimal solutions of $B_1$ and $B_2$ and obtain an optimal strategy for $B[i,j] \setminus c\ell(J)$. We need to consider every
possible prime interval $J$ in range $B[i,j]$ and take the minimum budget.

\begin{algorithm}[H]
\begin{algorithmic}[1]

\State \textbf{Input:} Bipartite permutation graph $G=(B,S,E)$ with ordering $<$ on vertices in $B,S$
i.e. $b_1 < b_2 < \dots < b_{n}, s_1  < s_2 <  \dots < s_{m}$\;
\State \textbf{Output:} Computing the budget for $G$ and optimal strategy\;

\For { $i=1$ to $i=n-1$ }

\For { $j=1$ to $j \le n-i$} 

 \State Let $H'=(B[j,j+i],N^*(B[j,j+i]))$ 
 
 \State Let $K'$ be the minimum number s.t. Optimal-Budget($H',K'$) is True. 
 
 \State Set $bg(H')=K'$ and let process of $H'$ be according to Optimal-Budget($H',K'$) 
 
  %\For {$j \le j' \le j+i$ s.t. $B[j',j'']$, $j+i+1 \le j'' \le |B|$ is a prime interval} 
  
   \State Let $H_r= H' \cup S_r$ ($S_r$ is the set of vertices who have neighbors in both $B[j+1,n],B[i,j]$)   

    \State Let $K'$ be the minimum number s.t. Optimal-Budget($H_r,K'$) is True. 
    
    \State Set $bg(H_r)=K'$ and let process of $H_r$ be according to Optimal-Budget($H_r,K'$)
    
 %\EndFor

  %\For {$j \le j' \le j+i$ s.t. $B[j'',j']$, $1 \le j'', j+i-1$ is a prime interval} 

   \State Let $H_l= H' \cup S_l$ ($S_l$ is the set of vertices who have neighbors in both $B[1,i-1],B[i,j]$)

    \State Let $K'$ be the minimum number s.t. Optimal-Budget($H_l,K'$) is True. 
    
    \State Set $bg(H_l)=K'$ and let process of $H_l$ be according to Optimal-Budget($H_l,K'$)
    
 %\EndFor

\EndFor

\EndFor 

\Statex 
\Function{Optimal-Budget}{$H=(B,S)$, $K$} 
\State \textbf{Input:} Bipartite permutation graph $H=(B,S,E)$ with ordering $<$ on vertices in $B,S$
%i.e. $b_1 < b_2 < \dots < b_{|B|}, s_1  < s_2 <  \dots < s_{|S|}$ and budget $K$\;
\State \textbf{Output:} Process $H$ with budget at most $K$, otherwise "False".\;
\If{$S= \emptyset$ and $K \ge 0$ OR $H$ is a bipartite clique and $|B|\leq K$}
\Return Process $H$\;
\EndIf
\If {there is a positive minimal subset $I$ with $bg(I) \le K$}
     process $I$ and $N^*(I)$
     
     \State \Return \textsc{BudgetPermutation} $(H[B \setminus I,S \setminus N^{*}(I)],K-\vert I\vert +\vert N^{*}(I)\vert)$\;
\EndIf

\If{ $|I| > K$ for all prime $I \subseteq B$}
\Return False\;
\EndIf
\If{a positive set $I$ with the smallest budget has $bg(I) > K$ }
\Return False\;
\EndIf

\For{ every prime interval $I$ of $H$ }
\State Let $H_1=(B_1,S_1)$ and $H_2=(B_2,S_2)$ where $B_1= \{b_1,b_2,\dots ,b_{i}\}$ and $B_2=\{b_{j},\dots,b_{n} \}$

 \hspace{2mm} $b_{i+1}$ is the first vertex of $c\ell(I)$ and $b_{j-1}$ is the last vertex of $c\ell(I)$ in the ordering $<$

\State Let $S_i$, $i=1,2$  be the set of vertices in $S$ that have neighbors in $B_i$
\Comment{$S_1 \cap S_2= \emptyset$}\;

\State Set Flag=Combine($H_1,H_2, K-|c\ell(I)|+|N^*(c\ell(I))|$).

 \If { Flag=True}

 \State Process of $H$ be $c\ell(I)$ together with process of $H \setminus ( c\ell(I) \cup N^*(c\ell(I)))$ by Combine

 \State return

 \EndIf
\EndFor

 \EndFunction
%\caption{Optimal-Budget}
%\label{func:opt-budget}

\end{algorithmic}
\caption{{\textsc{BudgetPermutation}} ( $H, K$)}
\label{alg:permutation}
\end{algorithm}

\begin{algorithm}[H]
\begin{algorithmic}[1]
\State \textbf{Input:} Optimal strategies for $H_1=(B_1,S_1,E_1), H_2=(B_2,S_2,E_2)$ and budget $K$\;
\State \textbf{Output:} "True" if we can process $H=H_1\cup H_2$ with budget at most $K$, otherwise "False".\;

\State Let $J_1$ be the first prime set in $H_1$ and $J_2$ be the first prime set in $H_2$.

\If {$|J_1| > K$ OR  $bg( H_2) > K- |c\ell(J_1)|+|N^*(c\ell(J_1))| $}
	\State Process $c\ell(J_2)$ and $N^*(c\ell(J_2))$
       \State Call $\textsc{Combine} (H_1, H_2 \setminus (c\ell(J_2) \cup N^*(c\ell(J_2))), K-|c\ell(J_2)|+|N^*	(c\ell(J_2))| ) $.

\Else{}  \State Process $ c\ell(J_1)$ and $N^*(c\ell(J_1))$,
         \State Call \textsc{Combine} $(H_1\setminus (c\ell(J_1) \cup N^*(c\ell(J_1))), H_2, K-|c\ell(J_1)|+|N^*(c\ell(J_1))| ) $.

\EndIf

\end{algorithmic}
\caption{{\textsc{Combine}} ( $H_1,H_2, K$)}
\label{combine}
\end{algorithm}

\begin{tm}\label{thm:poly-permutation}
Algorithm \ref{alg:permutation} solves the \textsc{Bipartite Ordering Problem} on a bipartite permutation graph with 
$n$ vertices in time $O(n^6 \log |B|)$. 
\end{tm}
\pf Let $H=(S,B,E)$ be a bipartite permutation graph with an ordering on its vertices
as described above. We use a dynamic programming table which keeps track of the the subgraph $H'$ induced by $B[i,j], N^*(B[i,j])$ and 
$b_i,b_{i+1},\dots,b_j$ is an interval in $B$. In the table we also keeps track of the subgraph $H''=(B'',S'')$ where $B''$ is a sub-interval of $B$
and $S''$ consists of vertices $N^*(B'')$ together with vertices of $S$ that are not initially in $N^*(B'')$ but are initially in 
$N^*(B'' \cup J)$ where $N(B'') \cap N(J)\neq \emptyset$ for some sub-interval $J$ of $B$. This instances appears after removing $c\ell(I)$ for some prime 
intervals $I$ of $B$. The number of such sub-instance is at most $O(n)$ for each interval $I$ of $B$. 

Now we show that Function Optimal-Budget correctly compute the budget for a given instance. 
The line 19 of the function is obvious. The correctness of lines 20-23 follow from Lemmas \ref{first-prime}, \ref{lm:minpos}, and 
Lemma \ref{correctness}.  

%We may assume that the optimal ordering for any graph $H'=(B',S')$ of $H$ where $B'$ is a subinterval of $B$ and $S'$ is a set of
%vertices in $S$ whose neighbors lie in $B'$ together with a prime interval that is either to the right of $B'$ or to the left of $B'$.
%Note that the number of such instances in polynomial of size of $H$.
%
%We start from the leftmost vertex $v$ in $B$ and move to the right, iteratively
%determining $bg$ in larger sub-graphs of $H$ from the right.
%We assume the optimal ordering for graph $H_1=H\setminus\lbrace v\cup N(v)\rbrace$ has
%been computed according to the rules of Algorithm \ref{alg:convex}.

We show how to find an optimal ordering for $H$ following the rules of Function Optimal-Budget.
First, we need to find all positive minimal sets. For bipartite permutation graphs, prime sets, the closure of a prime set ($c\ell(I)$), 
and any positive minimal set is an interval.  

%thus we must keep track of the budget needed
%to process intervals in $B$, and there are only polynomially many of these to
%consider (in particular, $O(n^2)$ where $B$ has $n$ vertices). Considering prime sets
%and closures of prime sets, when we consider an interval $J$ we need to consider $N^*(J)$ together
%with vertices of $S$ that are not initially in $N^*(J)$ but are initially in $N(J')$
%where $N(J) \cap N(J')\neq \emptyset$. This is because after processing $J'\setminus J$
%we may need to determine the budget for $J$ in the new instance and
%here the neighborhood of $N^*(J)$ contains new vertices. We need to keep track of $bg$ for the instances of this form, i.e. in 
%the Algorithm \ref{alg:permutation} the $B_1,B_2$ and they can be identified after removing $c\ell(I)$. 
Note that computing $c\ell(I)$ takes $O(n)$ and it is a straightforward procedure. Once $I$ is removed from $B$ there are two 
unique prime intervals (one on the right of $I$ and one in the left of $I$) that could potentially become positive and it can be added into $c\ell(I)$.  
% 
% This means that our dynamic programming table is based on those instances. 
% We note that 
% 
% 
% However, the number of
% these new intervals is at most $O(n^2)$. This task is in fact computing the budget for a
% proper sub-interval $J$ in $B$ together with the appropriate set of vertices in $S$
% which may need to be considered in $N^*(J)$. From now on we assume
% we can answer the queries such as $bg(c\ell(J)) \leq K$ for a given $w$ using a dynamic programming table.
% This allows us to be able to compute the optimal ordering for any graph $H'=(B',S')$  where $B'$ is a subinterval of $B$ and $S'$ is a set of
% vertices in $S$ whose neighbors lie in $B'$ together with a prime interval that is either to the right of $B'$ or to the left of $B'$.

When we consider processing a positive minimal set,  we not that according to Definition \ref{def:positive},
it does not have any proper positive subset. Therefore, it is the same as the case when we have
a bipartite permutation graph without any positive prime interval and no positive closure set
(Definition \ref{def:closure}).

%\begin{claim}
% Let $I$ be a prime interval in $B_i$ containing the last vertex of $B_i$ in ordering $\prec$ and let $K$ be the current budget.  Suppose there is no positive interval in $B_i$. Then there is an optimal strategy that $I$ is processed first if the following happens.
%
% \begin{itemize}
%  \item $bg (H_i \setminus (c\ell(I) \cup N^*(c\ell(I))) ) \le K-|c\ell(I)|+|N^*(c\ell(I))|$.
% \end{itemize}
%
%\end{claim}
Now suppose there is no positive prime interval. The optimal strategy starts with some prime interval $I$ 
and then it process the closure of that interval. After removing $c\ell(I)$ and $N^*( c\ell(I))$
we end up with two instances $H_1=(B_1,S_1)$ and $H_2=(S_2,B_2)$ where they are disjoint.
Note that no vertex $s \in S_2$ is adjacent to any vertex in $b \in B_1$ as otherwise the vertices in $N^*(c\ell(I))$ must be adjacent to 
$b$ (because of the min property of the min-max ordering $<$) which are not adjacent. No vertex $s' \in S_1$ is adjacent to any vertex $b' \in B_2$ as otherwise
the vertices in $N^*(c\ell(I))$ must be adjacent to $b'$ (because of the max property of min-max ordering) which are not adjacent. 

Now by similar argument as in the proof of Theorem \ref{thm:poly-TP-CB} we conclude that Combine
obtain an optimal strategy for $H_1 \cup H_2$, given the optimal strategy for $H_1$ and $H_2$.
Observe that in the algorithm we consider every possible interval $I$ therefore we obtain an optimal strategy to compute $bg(H)$.
For a prime set $J$, computing the $c\ell(J)$ takes $O(n)$. Combine algorithm takes $O(n^2)$ to obtain the strategy
for $H_1 \cup H_2$ (because at each steps it computes $c\ell(J_i)$, $i=1,2$ for the primes intervals in $H_1,H_2$).

%For each interval $I$ we call Function \ref{func:opt-budget}.  

For each interval $I$ of $B$ we call the Function Optimal-Budget at most three times. 
In Function Optimal-Budget we call the Combine function at most $O(|I|^2)$ times (there are at most $O(|I|^2)$ prime intervals). Therefore the running time of
Function Optimal-Budget is $O(|I|^4)$ and it is at most $O(n^4)$. There are at most $n^2$ intervals. 
Therefore the running time of Algorithm \ref{alg:permutation} is $O(n^6 \log|B|)$. The term $\log |B|$ is because we should binary search to obtain the optimial value $K'$ 
in lines 9,12.

\qed

\section{General Strategy}\label{sec:general-strategy}
It may not always be the case that all positive sets can be identified
in polynomial time. But, if positive sets can be identified, the
following is a general strategy for processing an input bipartite graph $H$
and given budget $K$.

1. If there exist positive sets in $B$, process a positive minimal
set $I$, set $H = (B\setminus I, S\setminus N^*(I))$, update $K$ to  $K-|I|+|N^*(I)|$ and repeat step 1.

2. If no positive set exists, choose in some way the next prime
set $I$ to process, set $H = (B\setminus I, S\setminus N^*(I))$, update $K$ to be $K-|I|+|N^*(I)|$ and go to step 1.

Note that each time a prime set $I$ is processed, we end up processing
$c\ell(I)$.  Even if we can identify the prime and positive sets, it
remains to determine in the second step the method for choosing the next
prime to process.  We address this issue and give the full algorithm and
proof for Theorem \ref{thm:main} in the next subsection.  Note that Lemma
\ref{lm:minpos} implies that without loss of generality we can assume that
when a prime set $I$ is processed the remainder of $c\ell(I)$ is processed
next, as it is stated in the next corollary.

\begin{cor}
  Let $H = (B, S)$ be a bipartite graph that can be processed with budget
  at most $K$ with an ordering that processes prime set $I$ first. Then
  there is a strategy for $H$ that processes $c\ell(I)$ first and uses budget
  at most $K$.
\end{cor}
% Jeff 13 Jul 15 - added the corollary as explicitly stated, with no proof,
%  since it is ``obvious'' given the lemma above.

\subsection{Algorithm and Correctness of Proof for Theorem \ref{thm:main}}\label{sec:main-pf}

In this section we give the algorithm and proof for Theorem \ref{thm:main}, that
we can solve the bipartite graph ordering problem for some classes of graphs. From the previous section it remains to determine how to choose
which prime set to process first when there are no positive sets that can be
processed.

\begin{dfn} \label{dfn:pre-order}
Let $I,J$ be prime subsets. We say that $I$ is {\em potentially after} $J$ for current budget $K$ if

\begin{enumerate}

 \item $|I|  > K$, or
 %\vspace{2mm}
 %\item $|N^*( c\ell(I))|-|c\ell(I)|-bg(c\ell(J) \setminus c\ell(I))  < |N^*(c\ell(J))|-|c\ell(J)|- bg(c\ell(I) \setminus c\ell(J))$
\item
  $bg(c\ell(J)\setminus c\ell(I)) > K - |c\ell(I)| + |N^*(c\ell(I))|$
\end{enumerate}
% Jeff 13 Jul 15 - mentioned current budget in definition, and updated to match definition 10.
\end{dfn}

%Note that {\em potentially before } defines a total ordering on all the prime subsets.
%Note that if $c\ell(I) \cap J = \emptyset$ then instead of $bg(c\ell(J) \setminus c\ell(I))$ we have $|c\ellJ|$.

%Definition \ref{dfn:pre-order} is a first attempt at choosing which
%prime set to process first; the idea is to process prime sets which
%can be processed first and which would result in a higher
%budget if processed first.  The intuition behind this definition is that  if $I$ is {\em first} according to Definition \ref{dfn:pre-order}
%but in some optimal process $opt$, $J$ is processed before $I$ then we would be able to exchange $I$ and $J$ in the $opt$. For clarification we have singled out this case in the proof of the correctness of the algorithm. However, this may not  be the case as when $opt$ processes $J$ first, $I$ may no longer be the first prime subset according to Definition \ref{dfn:pre-order} and hence Definition \ref{dfn:pre-order}  is not quite enough.  We will see that the recursive Definition \ref{dfn:order} gives the right order in which to consider prime sets.
Definition \ref{dfn:pre-order} is a first attempt at choosing which
prime set to process first. The idea is to consider whether it is
possible to process $I$ before $J$.  Item 2 in the definition states
that $J$ could not be processed immediately after $I$.  However, this
formula is not sufficient in general because we must consider
orderings that do not process $I$ and $J$ consecutively, and we must
take into account that for whatever ranking we define on the prime
sets the ranking may change as the algorithm processes prime sets.
For clarification we have singled out the case when $I$ and $J$ are
processed consecutively in the proof of correctness of the algorithm.

If two prime sets $I$ and $J$ are not processed consecutively by the
$opt$ strategy, we should adapt Item 2 of Definition
\ref{dfn:pre-order} to take into account all vertices that would be
processed in between by our algorithm.  We call this set of vertices
the ``Superset'' of $J$ with respect to $I$, defined precisely by the recursive
Definitions \ref{dfn:superset} and \ref{dfn:order}.
% Jeff 13 Jul 15 - updated the text to hopefully more clearly explain
% where the definitions are coming from.

\begin{dfn} \label{dfn:superset}
% JJK, interval - what does this mean?  do we mean prime set? OK prime set
% JJK, is this definition circular, this definition refers to the next one, which refers back to this one? Yes it is circular

%%% Akbar, 2/2/2016
 Let $I$ and $J$ be two prime subsets. For current budget $K$, the \emph{Superset of $J$ with respect to $I$, denoted as $Superset_{I}(J)$,} is defined as follows.
$Superset_{I}(J)$ contains $c\ell(J)$ and at each step a set $c\ell(J_i)$ is added into $Superset_{I}(J)$ from $B \setminus Superset_{I}(J)$ where $J_i$ is first according to the lexicographical order of prime sets such that no prime set is before  $J_i$  according to the ordering in Definition \ref{dfn:order}.
We stop once $c\ell(I)$ lies in $Superset_{I}(J)$.

% Consider prime set $J_i$, $i \ge 1$ such that $J_i$ does not potentially come before $I_{i-1}$ in $H_{i-1}$.
 %If $J_i$ potentially comes before $I_{i-1}$ in $H_i=H_{i-1} \setminus (c\ell(J_{i-1}) \cup N^*(c\ell(J_{i-1}))$ and
 %and is the first according to potentially before relation in $H_{i-1}$  then add $c\ell(J_i)$ to $Superset_{I}(J)$ and set $I_i=I_{i-1} \setminus c\ell(J_i)$.
\end{dfn}

%Superset of $J$ with respect to $I$, is the set containing $c\ell(J)$ and sets $c\ell(J_i)$ of vertices that are processed
%after processing $c\ell(J)$ according to the Definition \ref{dfn:order} until the entire set $I$ is processed.

\begin{dfn}\label{dfn:order}
For current budget $K$, we say prime subset $I$ is {\em after} prime subset $J$ if
\begin{enumerate}
  %\item $I,J$  are siblings $|N^*(I)| > |N^*(J)|$.
  \item $|I| > K$, or
  %\vspace{1mm}

 % \item $|N^*( c\ell(I))|-|c\ell(I)|-bg(J \setminus c\ell(I))  \ge |N^*(c\ell(J))|-|c\ell(J)|- bg(I \setminus c\ell(J))$
 % \vspace{1mm}
  \item  $bg( Superset_{I}(J) \setminus c\ell(I)) > K- |c\ell(I)|+|N^*(c\ell(I))|$

 \end{enumerate}
\end{dfn}

Definition \ref{dfn:order} states that $I$ is after $J$ if it is too
large for the current budget (Item 1) or cannot be processed before
$J$ using the ordering implied by Definitions \ref{dfn:superset} and
\ref{dfn:order} (Item 2).
% Jeff 13 Jul 15 - say in words what the ordering means.
Note that if $I$ is processed right after $c\ell(J)$ then Item 2 in Definition \ref{dfn:order} agrees with
Definition \ref{dfn:pre-order}. In the graph induced by $\{J,J_1,J_2,I,D,E,F,L,P\}$ in Figure \ref{fig:2}, we have $Superset_{I}(J)=J \cup J_1 \cup J_2$ with respect to any current budget $K$ at least 12.

We point out that Definitions \ref{dfn:superset} and \ref{dfn:order} are
recursive, and a naive computation of the ranking would not be efficient.
We describe how to efficiently compute the ranking for the classes of graphs of Theorem \ref{thm:main} using
dynamic programming in Sections \ref{sec:poly-TP-CB} and \ref{sec:poly-permutation}. The main description of our general strategy is given in Algorithm \ref{fig:alg}.

\begin{algorithm}[H]
\begin{algorithmic}[1]
\State \textbf{Input:} $H=(B,S,E)$ and budget $K$\;
\State \textbf{Output:} "True" if we can process $H$ with budget at most $K$, "False" otherwise.\;
\If {$S= \emptyset$ and $K \ge 0$}\label{Alg-general-firstIF}
 \Return True\;
\EndIf \label{Alg-gnr-end-firstif}
\If{ $|I| > K$ for all prime $I \subseteq B$} \label{Alg-general-secondIF}
\Return False\;
\EndIf \label{Alg-gnr-end-secondif}
\If {there is a positive minimal subset $I$ with $bg(I) \le K$} \label{Alg-general-thirdIF}
     \Return \textbf{Budget} $(H[B \setminus I,S \setminus N^{*}(I)],
K-\vert I\vert +\vert N^{*}(I)\vert)$\;
\EndIf \label{Alg-gnr-end-thirdif}
\If{a positive set $I$ with the smallest budget has $bg(I) > K$ }  \label{Alg-general-fourthIF}
\Return False\;
\EndIf \label{Alg-gnr-end-fourthif}
\State Let $I$ be the lexicographically first prime subset with no other prime set before it according to ordering in Definition \ref{dfn:order}.

\If {no such $I$ exists} \label{Alg-general-fifthIF}
    \Return False
\Else {}
     \Return \textbf{Budget} $(H[B \setminus I,S \setminus N^{*}(I)], K-\vert I \vert +\vert N^{*}(I)\vert)$\;
\EndIf \label{Alg-gnr-end-fifthif}
\end{algorithmic}
\caption{{\textbf Budget } ( $H=(B,S,E)$, $K$)}
\label{fig:alg}
\end{algorithm}

The algorithm determines whether $bg(H) \leq K$. Note that the exact
optimal value can be obtained by using binary search, and since the
optimal value is somewhere between 0 and $|B|$ the exact computation
is polynomial if the decision problem is polynomial.

Before considering the running time for the graph classes of Theorem \ref{thm:main} we
first demonstrate that the algorithm in Algorithm \ref{fig:alg} decides correctly,
though possibly in exponential time, for any instance $( H = (B, S), K)$.

\begin{lemma}\label{lm:optimal}
  For any $K$ and bipartite graph $H$, the Budget algorithm  (Algorithm \ref{fig:alg}) correctly
  decides if $bg(H) \leq K$ or not.
\end{lemma}
\pf%\pfB{Proof of Lemma \ref{lm:optimal}}
We show that if $bg(H)=K$ then there exists an optimal solution $opt'$ with
budget $K$ in which subset $I$ as described in the algorithm is processed
first. We use induction on the size of $B$, meaning we assume that for smaller instances,
there is an optimal process that considers the prime subsets according to the rules of our algorithm.

Correctness of Lines 3 is clear and the correctness of line 4 follows from Lemma \ref{first-prime}.
The correctness of steps 5 follows from Lemma \ref{lm:minpos}. Suppose Line 6 were incorrect. Then
all positive subsets would have budget above $K$. Let $I^+$ be one such subset and yet if 
Line were incorrect there would be a way to process $I^+$ with budget at most $K$ in $H$. In that case, we would process some negative set
$I^-$ which somehow reduces the budget of processing $I^+$; this can
only be so if $I^-\cap I^+ \neq \emptyset$.  In this case the Lemma
\ref{correctness} states that $I^+ \cup I^-$
is itself a positive set with budget at most $K$, a contradiction to
the premise of step 6.
%
% \begin{claim}\label{correctness}
% Suppose that $I^{+}$ is a positive subset with $bg(I^+) > K$
% and $I^{-}$ is a negative subset where $bg(I^{-}) \le K$ and $I^+ \cap I^- \ne\emptyset$.
% If $bg( I^{+} \cup I^{-}) \le K$ then $I^{+} \cup I^{-}$ forms a positive subset.
% \end{claim}
% \pf Let $X=I^-\cap I^+$. By the assumption that $I^+$ can be processed after processing
% $I^-$ we have $bg(I^+\setminus X)\leq K-|I^-|+|N^*(I^-)|$. On the other hand, since $bg(I^+)>K$
% then $K-|X|+|N^*(X)|<bg(I^+\setminus X)$. From these two we can conclude that:\\
% \begin{equation}\label{eq}
% |N^*(I^-)|>|I^-|-|X|
% \end{equation}
%
% Moreover, because $I^+$ is a positive set then $|N^*(I^+)|\geq |I^+|$.
% By (\ref{eq}), $I^+$ being positive, and the fact that
% $|N^*(S \cup T)| \geq |N^*(S)| + |N^*(T)|$ for any $S$ and $T$, we have
% $|N^*(I^+\cup I^-)| \geq |N^*(I^+)|+|N^*(I^-)|\geq |I^+|+|I^-|-|X| = |I^+\cup I^-|$, i.e., $I^+\cup I^-$ is a positive subset.
% \qed \\

We are left to verify Lines 7-9, so we continue by assuming there are
no positive subsets. Let $I$ be the first prime set according to Definition \ref{dfn:order}.
Suppose for the sake of contradiction that the optimal solution $opt$ processes
prime subset $J$ before $I$. In what follows we show that we can modify $opt$ and process $I$ as the first prime set. % where $I$ is the first prime set according to Definition \ref{dfn:order} (otherwise we can use induction to say the
%algorithm is correct).
Note that, since there is no positive subset at the beginning, $opt$ processes $c\ell(J) \setminus J$ after $J$.\\
%% Suppose $I$ is the first subset according to the rules of the
%% algorithm.  Note that $I$ would also be the first considered within $H
%% \setminus (c\ell(J))$, and by induction this would be correct for
%% $H\setminus (c\ell(J))$.

%{\textbf Case 1.}

Suppose that by induction hypothesis (rules of our algorithm) the $opt$
would place $I \setminus c\ell(J)$ first in $H'=(B \setminus c\ell(J), S\setminus N^*(c\ell(J)))$. In this case $Superset_I(J) \setminus c\ell(I)$
is just $c\ell(J) \setminus c\ell(I)$, and in this case Definitions \ref{dfn:pre-order} and
\ref{dfn:order} coincide.
% Jeff 13 Jul 15 - point out that this the case where the definitions coincide.

We show that we can modify $opt$ to process $c \ell(I)$ first and then
$J \setminus c\ell(I)$ next while still using budget at most $K$. Suppose this is not the
case. Now we have the following

\begin{itemize}
%\item[(a)] $K- |c\ell(J)|+|N^*(c\ell(J))|  \ge bg(c\ell(I) \setminus c\ell(J)) $

\item [(a)] $bg(c\ell(J) \setminus c\ell(I))  > K- |c\ell(I)|+|N^*(c\ell(I)) |$

\end{itemize}

%The inequality (a) follows by $opt$ being an ordering with budget at most $K$ that
%processes $c\ell(J)$ first: since $opt$ processes $c\ell(I) \setminus c\ell(J)$ after $c\ell(J)$, the budget must be at least
%$bg(c\ell(I) \setminus c\ell(J))$  after processing $c\ell(J)$.

The inequality (a) follows from the assumption that we cannot process $c\ell(I)$ first and then immediately processing
$ J \setminus c\ell(I)$. However, this is a contradiction to the fact that $I$ is before $J$ according to Definition \ref{dfn:order}.

We also note that since $bg( c\ell(J))  \le bg( Superset_{I}(J) \setminus c\ell(I) ) \le K- | c\ell(I)|+|N^*(c\ell(I))|$, we can also process
the entire $c\ell(J)$ after processing $c\ell(I)$. Therefore we can exchange processing $c\ell(I)$ with $c\ell(J)$ and follow the
$opt$ in the remaining. \\

%{\textbf Case 2.}

We are left with the case that $c\ell(J)$ is processed first by $opt$, and the rules
of the algorithm (second item in Definition \ref{dfn:order}) would process some prime subset $L$ different from $I \setminus c\ell(J)$ next.
This would imply that there is some prime subset $L$ that is considered before the last remaining part of $I$ in $B \setminus c\ell(J)$.
By induction hypothesis we may assume that the $opt$ processes the prime subsets according to the second item in Definition \ref{dfn:order}.
These would imply $L$ is in $Superset_{I}(J)$.  At some point $I$ or the remaining part of $I$ becomes the first set to process according to
the rules of the algorithm and this happens at the last step of the definition of  $Superset_{I}(J)$.
However, since there is no other prime subset before $I$ according to Definition \ref{dfn:order} we have
$bg( Superset_{I}(J) \setminus  c\ell(I)) \le K- |c\ell(I)|+|N^*( c\ell(I))|$. Therefore  we can process  $c\ell(I)$ first and next
$c\ell(J) \setminus c\ell(J)$ and then follow $opt$.

It remains to show that if $bg(H)\leq K$ then there exists a prime subset $I$
that is the lexicographically first prime subset with no other
prime set before it according to the ordering in Definition \ref{dfn:order}.
Suppose there exists an ordering for $H$ with budget at most
$K$ as follows: $c\ell(J), c\ell(J_1),\dots , c\ell(J_r)$.
By induction assume that the Budget Algorithm returns ``true'' for
instance $H\setminus \lbrace c\ell(J)\cup N^*(c\ell(J))\rbrace$ with
budget $K-|c\ell(J)|+ |N^*(c\ell(J))|$ and the output ordering is
$c\ell(J_1),\dots , c\ell(J_r)$. Therefore, by Definition
\ref{dfn:superset}, $Superset_{J_i}(J)=\cup_{t=1}^{i}c\ell(J_t)$ for
$1\leq i\leq r$. Observe that $J$ is not after any prime subset
by Definition \ref{dfn:order} which leads us to have $J$ as a valid
``first'' prime subset in $H$ for the algorithm.
\qed \\

%Note that Lemma \ref{lm:optimal} implies that the algorithm
%can be used to find the optimal value for \emph{any} graph, while Theorems
%\ref{thm:poly-convex} and \ref{thm:poly-TP-CB} show that the algorithm is efficient for the classes of
%graphs mentioned in Theorem \ref{thm:main}.
A naive implementation of the algorithm would consider all possible
orderings of prime sets to determine the ordering in step 4 of the algorithm, and
in the worst-case an exponential number of sets may need to be considered to
identify the prime and positive minimal sets.  A careful analysis can be taken
to show that the running time of the algorithm in the general case is
exponential.  In the next two sections we show that for the graph classes
of Theorem \ref{thm:main} the running time is polynomial.
% Jeff 13 Jul 15 - mention here that the algorithm is general, and then in the
%  last section ask the question whether the algorithm can be used for a
%  dichotomy theorem.

%
%Theorem \ref{thm:main} follows from the following two lemmas.
%Lemma \ref{lm:optimal} states that our algorithm is correct, and this
%holds for all graphs.

%%%%%%%%%%%%%%%%%%%%%%%%%%%%%%%%%%%%%%%%%%%%%%%%%%%%%%%%%%%%%%%

\section{Future Work and Open Problems} \label{sec:future}

We have defined a new scheduling or ordering problem that is
natural and can be used to model processes with precedence
constraints.  As with any optimization problem there are many avenues
of attack.  In this work we have focused on determining for which
classes of graphs the bipartite graph ordering problem can be solved in
polynomial time.  Our ultimate goal in this direction is a dichotomy
classification of polynomial cases and NP-complete cases.  The algorithm
in the proof of Theorem \ref{thm:main} finds the optimal budget for
all graphs $H$, and the algorithm was shown to run in polynomial time
for the classes of graphs mentioned in Theorem \ref{thm:main}.  We pose the
question whether the algorithm can be the basis of a dichotomy theorem:
are there classes of graphs which can be solved in polynomial time but
for which our algorithm does not run in polynomial time?
% Jeff 13 Jul 15 - modified the last sentences a bit to ask a specific
%  question.
%
%  We could also ask what the worst-case running time is of our
%  algorithm: if it is linear exponential, we could use it for the
%  exponential time algorithm instead of the reduction to previous work.
%  But that doesn't seem worth mentioning here, it's just something
%  we should think about.

As with all optimization problems the bipartite graph ordering problem can
also be studied from a number of other angles, including approximation
and hardness of approximation, fixed parameter algorithms, and faster
exponential-time algorithms.
%We have not spent much effort in
%considering these other directions and hope they can be fruitful areas
%for the community to consider.
A particular graph class to consider in
each of these areas is that of circle bipartite graphs, because these graphs are
of particular interest in the application to molecular folding \cite{GFWT08,MH98,TMRM09}.
%
% Our techniques are designed for ordering \emph{bipartite} directed
% graphs.  New techniques may be needed to make progress on solving the
% ordering problem with precedence constraints and budget minimization
% on \emph{non-bipartite} directed acyclic graphs.

\paragraph{\textbf Acknowledgments}
We would like to thank
Pavol Hell, Ladislav Stacho, Jozef Hale\u{s}, Cedric Chauve and Geoffrey Exoo for many useful
discussions.

\vspace{-3mm}

% Jeff

\end{document}